\documentclass[twocolumn,superscriptaddress,preprintnumbers,amsmath,10pt,aps,prl,nobibnotes,longbibliography]{revtex4-1} 

\usepackage{algorithm, algorithmic}

\usepackage{amsmath}
\usepackage{amsfonts}
\usepackage{bm}
\usepackage{color}
\usepackage{verbatim}
\usepackage{graphicx}
\usepackage{lipsum}
\usepackage[utf8]{inputenc}
\usepackage{times}
\usepackage{graphicx}
\usepackage{bm}
\usepackage{amssymb,multirow}
\usepackage[dvipsnames]{xcolor}
\usepackage{soul}
\usepackage[normalem]{ulem} 
\usepackage[mathlines]{lineno}
\usepackage{bbm}
\usepackage{siunitx}
\usepackage{array}
\usepackage{physics}
\usepackage{hyperref}
\hypersetup{
    colorlinks=true,
    linkcolor=blue,
    citecolor=blue,
    urlcolor=blue
    }

\def\Veca{\mathbf{a}}
\def\Vecb{\mathbf{b}}
\def\Vecc{\mathbf{c}}

\def\Vech{\mathbf{h}}

\def\Vecu{\mathbf{u}}
\def\Vecv{\mathbf{v}}

\def\VecD{\mathbf{D}}

\def\VecM{\mathbf{M}}
\def\VecP{\mathbf{P}}

\def\VecS{\mathbf{S}}
\def\VecT{\mathbf{T}}
\def\VecU{\mathbf{U}}
\def\VecV{\mathbf{V}}

\begin{document}

\title{All-optically untangling light propagation through multimode fibres}

\author{Hlib Kupianskyi}
\email{hk422@exeter.ac.uk}
\affiliation{Physics and Astronomy, University of Exeter, Exeter, EX4 4QL. UK.}
\author{Simon A.~R.~Horsley}
\affiliation{Physics and Astronomy, University of Exeter, Exeter, EX4 4QL. UK.}
\author{David~B.~Phillips}
\email{d.phillips@exeter.ac.uk}
\affiliation{Physics and Astronomy, University of Exeter, Exeter, EX4 4QL. UK.}

\begin{abstract}
When light propagates through a complex medium, such as a multimode optical fibre (MMF), the spatial information it carries is scrambled. In this work we experimentally demonstrate an {\it all-optical} strategy to unscramble this light again. We first create a digital model capturing the way light has been scattered, and then use this model to inverse-design and build a complementary optical system -- which we call an {\it optical inverter} -- that reverses this scattering process. Our implementation of this concept is based on multi-plane light conversion, and can also be understood as a diffractive artificial neural network or a physical matrix pre-conditioner. We present three design strategies allowing different aspects of device performance to be prioritised. We experimentally demonstrate a prototype optical inverter capable of simultaneously unscrambling up to 30 spatial modes that have propagated through a 1\,m long MMF, and show how this enables near instantaneous incoherent imaging, without the need for any beam scanning or computational processing. We also demonstrate the reconfigurable nature of this prototype, allowing it to adapt and deliver a new optical transformation if the MMF it is matched to changes configuration. Our work represents a first step towards a new way to see through scattering media. Beyond imaging, this concept may also have applications to the fields of optical communications, optical computing and quantum photonics.
\end{abstract}

\maketitle

\noindent As their name suggests, multimode optical fibres (MMFs) support the transmission of multiple spatial modes, recognisable as unique patterns imprinted in the electric field of guided laser light~\cite{snyder2012optical}. These spatial modes are capable of acting as independent information channels, offering the tantalising prospect of ultra-high density information and image transmission through hair-thin strands of optical fibre~\cite{rademacher2021peta,cao2023controlling}. Such technology has a wealth of applications, from high-resolution micro-endoscopy deep inside the body~\cite{gigan2022roadmap,stibuurek2023110,wen2023single}, to space-division multiplexing through short-range optical interconnects in data centres~\cite{richardson2013space,cristiani2022roadmap}, and emerging forms of quantum communication and photonic computing~\cite{rahmani2022learning,teugin2021scalable,leedumrongwatthanakun2020programmable}.
However, there are significant challenges to overcome before the high data capacity of MMFs can be fully unlocked.

An optical field illuminating one end of a MMF typically emerges from the other end unrecognisably spatially scrambled -- a consequence of modal dispersion and cross-talk. This presents a major hurdle to spatial signal transmission and imaging through MMFs, as the light must somehow be unscrambled again to recover images and data~\cite{cao2023controlling}. A number of techniques to achieve this are currently under development. A widely applicable strategy involves first creating a digital model of the way the fibre scrambles light. This can be accomplished by measuring the fibre's transmission matrix (TM) -- a linear operator encapsulating how any spatially coherent optical field will be transformed upon propagation through the MMF~\cite{popoff2010measuring}. Once the TM is known, it links monochromatic fields at either end of the MMF, and so knowledge of the field at one end enables computational recovery of the field at the other end~\cite{choi2012scanner,lee2022confocal} -- a technique closely related to coherent optical multiple-input multiple-output (MIMO) in the optical communications domain~\cite{shah2005coherent,rademacher20221}. 
\begin{figure}[t]
   \includegraphics[width=0.45\textwidth]{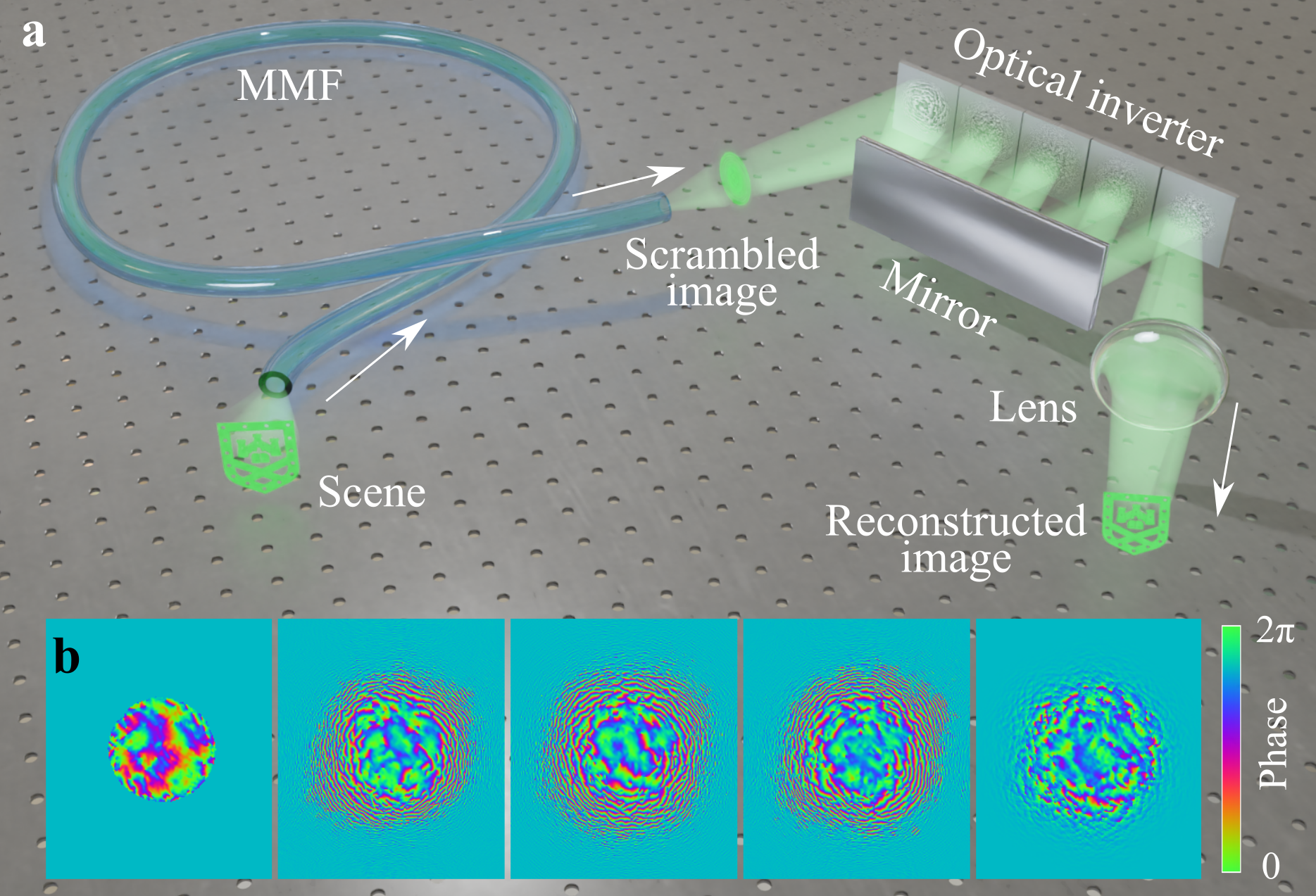}
   \caption{{\bf Optical inversion concept}. (a) Light emanating from a scene propagates through an MMF, which spatially scrambles the optical field. The light subsequently propagates through an optical inverter, which performs a passive unscrambling operation. Thus, an image of the scene at the fibre input is reconstructed at the inverter output. The optical inverter consists of a cascade of diffractive optical elements that are inverse-designed to match the transmission properties of the MMF. (b) An example of the phase profiles of a 5-plane optical inverter implemented experimentally.}
   \label{Fig:concept}
\end{figure}

Knowledge of the TM also enables scanning imaging through MMFs to be accomplished~\cite{vcivzmar2012exploiting}. This method, known as wavefront shaping~\cite{vellekoop2007focusing}, uses a spatial light modulator (SLM) to dynamically structure input optical fields so they transform into focused spots after propagation through the fibre~\cite{papadopoulos2012focusing}. By scanning a spot over the scene, and recording the total return signal that has emanated from each of the known spot locations, reflectance or fluorescence images can be reconstructed. Wavefront shaping is a powerful technique that has enabled a wide variety imaging modalities through MMF-based micro-endoscopes~\cite{loterie2015digital,gusachenko2017raman,turtaev2018high,ohayon2018minimally,tragaardh2019label,cifuentes2021polarization,leite2021observing,stellinga2021time}. However, in these methods the spatial information is essentially unscrambled {\it one mode at a time} -- which severely limits imaging frame-rates, and is not compatible with wide-field or super-resolution imaging~\cite{betzig2006imaging,rust2006sub} through fibres.

To take full advantage of the parallel information channels supported by MMFs, we would ideally be able to disentangle all propagating spatial modes {\it simultaneously}, to a high fidelity, and with minimal computational overhead~\cite{liu20221,doerr2011proposed,tanomura2020monolithic}. In this article, we show how the unscrambling operation can be achieved passively in an all-optical manner, with a latency limited only by the speed of light. Armed with knowledge of a fibre's TM, we design a complementary optical system -- crafted through the process of inverse design -- that reverses the scattering process imparted by the MMF. We refer to this device as an {\it optical inverter}~\cite{butaite2022build}. It brings closer the vision of being able to simply look through an optical fibre to directly see the scene at the other end.

An optical inverter must precisely manipulate many spatial modes simultaneously. Realisation of photonic systems capable of on-demand high-dimensional spatial mode transformation is a challenging task, with techniques still in their infancy~\cite{molesky2018inverse,bogaerts2020programmable,resisi2020wavefront}. Despite their high resolution, a single reflection from a two-dimensional SLM or metasurface cannot achieve an arbitrary multi-modal transformation -- for this the interaction of light with a three-dimensional photonic architecture is required~\cite{kupianskyi2023high}. Our concept, shown in Fig.~\ref{Fig:concept}, relies on a technology known as {\it multi-plane light conversion}~\cite{labroille2014efficient}, and can also be understood as a physically realised {\it diffractive artificial neural network}~\cite{lin2018all}. Light emerging from the MMF reflects from a cascade of specially designed diffractive optical elements -- here referred to as `phase planes' (see Fig.~\ref{Fig:concept}(b)), each separated by free-space. These static phase planes successively rearrange the spatial information carried by the light, operating on all modes simultaneously, and enacting the inverse transformation to that applied by the fibre itself. After this process, images of input optical fields are formed at the output of the inverter, without the need for any computational processing.

\begin{figure*}[t]
   \includegraphics[width=0.8\textwidth]{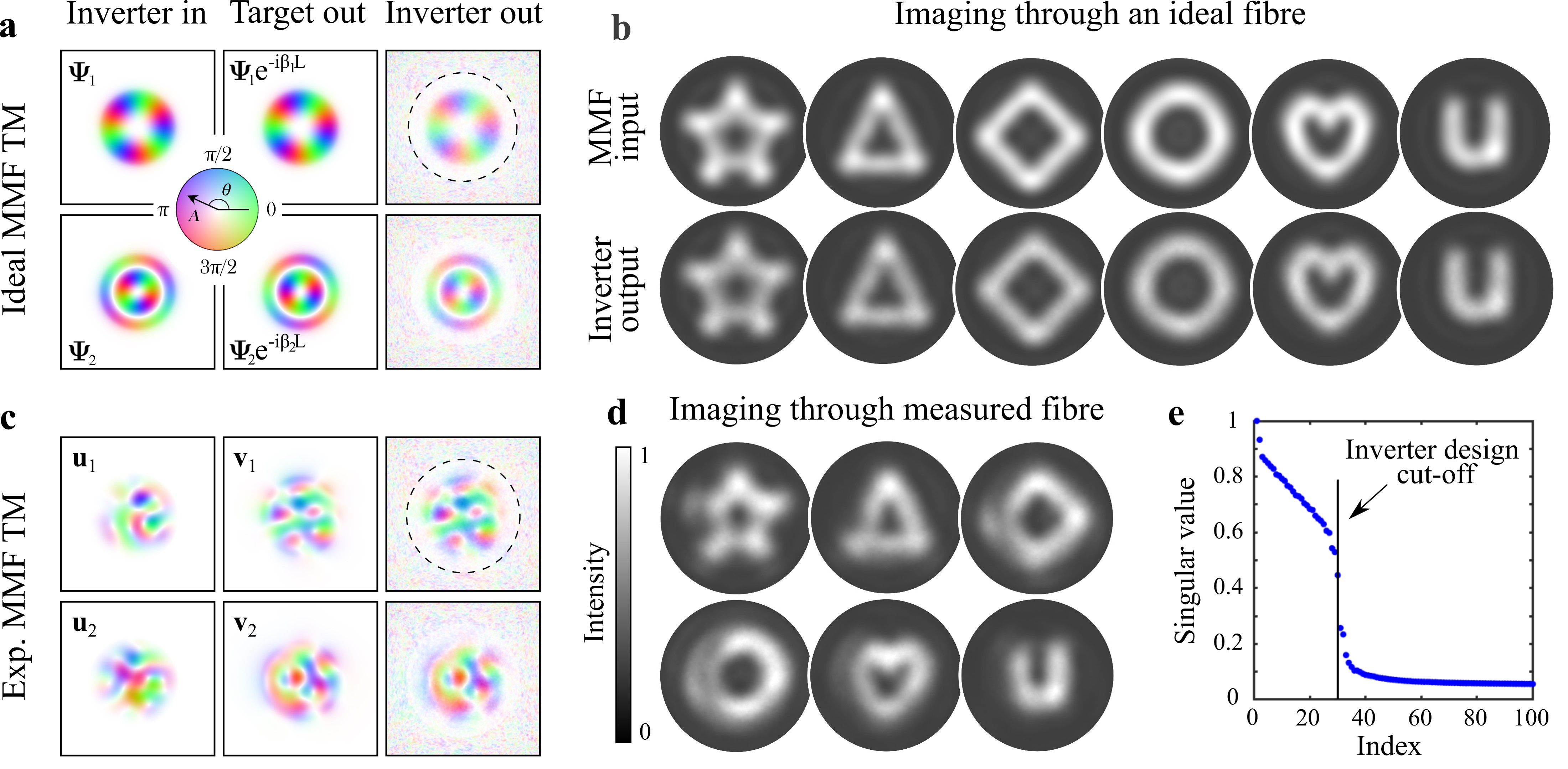}
   \caption{{\bf Simulations of optical inverter performance} (a,b) Eigenmode-based inverter coupled to an ideal fibre (unscrambling 42 modes using 5 phase planes of resolution 700$\times$700 pixels). (a) Two examples (simulations) of PIMs imparted a global phase-shift by the inverter. Left column: input PIMs. Middle column: target phase-shifted output PIMs. Right column: simulated output from MPLC inverter. The target output modes are projected into a disk (black hatched circle) with a mean fidelity of $f=97\,\%$ and efficiency of $e=12\,\%$ (see Methods for definitions) and uncontrolled light is scattered outside this disk. See SI \S 4 for a visualisation of the transformation of all eigenmodes. (b) Examples (simulations) of incoherent images transmitted through the ideal MMF-inverter system. Upper row: Incoherent fields incident on MMF input facet. Lower row: Corresponding incoherent fields re-imaged at the output of eigenmode-based inverter. (c-e) SVD-based inverter coupled to a realistic fibre (unscrambling 30 modes using 5 planes of resolution 700$\times$700 pixels). The fibre TM was measured experimentally. (c) Two examples (simulations) of left-hand singular vectors $\Vecu_n$ transformed into corresponding right-hand singular vectors $\Vecv_n$ by the inverter. Left column: input left-hand singular vectors. Middle column: target output right-hand singular vectors. Right column: simulated output from MPLC inverter. See SI \S 5 for a visualisation of the transformation of all modes. (d) Incoherent fields re-imaged at the output of the SVD-based inverter (simulations). (e) Singular value distribution of the experimentally measured fibre TM (top 100 singular values shown). Vertical line shows the SVD-inverter design cut-off value at $N=30$.}
   \label{Fig:svd_inv}
\end{figure*}

We recently proposed this optical inversion concept, as a way to unscramble light propagation through MMFs, in a numerical study~\cite{butaite2022build}. Here we experimentally implement a prototype optical inverter, with its design tuned to a specific MMF supporting up to $\sim$30 spatial modes. We also demonstrate how this prototype inverter can be adapted to match a new fibre TM if the bend configuration of the MMF is perturbed -- pointing towards future applications involving flexible fibres. While our work predominantly targets imaging applications, the concepts we introduce here may also prove fruitful in the fields of optical communications, optical computing and quantum photonics. \\

\noindent{\large{\bf Optical inverter design}}\\
We design an optical inverter complementary to a short length (1\,m) of step-index MMF with a core diameter of $d=25\,\mu$m, and numerical aperture $\text{NA}=0.1$. At a wavelength of ${\lambda = 633}$\,nm, this MMF supports $N = 42$ spatial modes per polarisation channel (given by $N\sim\left(\pi d\text{NA}/2\lambda\right)^2$). We demonstrate three inverse design protocols that enable different performance criteria to be optimised.\\

\noindent{\bf Eigenmode-based inverter design}: We start by considering how to unscramble light transmitted through an ideal MMF. Under the weakly guiding approximation (i.e.\ low $\text{NA}$), the MMF eigenmodes are a set of $N$ circularly polarised {\it propagation invariant modes} (PIMs)~\cite{ploschner2015seeing}.  The PIMs maintain an unchanging transverse field profile as they propagate along the fibre, denoted by $\boldsymbol{\Psi}_n$, where $n$ indexes the mode: ${n\in\{1,2,3,...,N\}}$ -- see Fig.~\ref{Fig:svd_inv}(a) and Supplementary Information (SI) \S 4 for examples. Being eigenmodes of the system, the ideal PIMs exhibit negligible mode-dependent loss and modal coupling during propagation.

Each PIM has a mode-dependent propagation constant $\beta$ -- describing the rate at which its global phase shifts as it propagates along the fibre. Consequently, mode $n$ picks up a mode-dependent global phase delay of ${\beta_n L}$ upon reaching the output of a fibre of length $L$. This spatial mode dispersion causes the interference pattern cast by a superposition of PIMs to be different at the input and output fibre facets, resulting in the observed spatial scrambling of optical fields. Therefore, in matrix form the TM of an ideal MMF, in a single circularly-handed polarisation state and represented in real-space (pixellated) input and output bases, is well-approximated by ${\VecT_{\text{fib}} = \VecP\VecD\VecP^\dagger}$. Here $\VecP$ transforms from fibre mode space to real-space, and $\VecD$ is a diagonal unitary matrix encoding the phase delay accumulated by the PIMs along its diagonal. This TM links the arbitrary vectorised input field $\Vecu$ to the corresponding output field $\Vecv$ via: ${\Vecv = \VecT_{\text{fib}}\Vecu}$.

In this ideal case, the task of the optical inverter is to reverse these mode-dependent phase delays: it should enact the transform ${\boldsymbol{\Psi}_n\rightarrow\boldsymbol{\Psi}_n\exp(-i\beta_n L)}$ for all $N$ PIMs simultaneously.
Therefore, the TM of the optical inverter is given by ${\VecT_{\text{inv}} = \VecP\VecD^\dagger\VecP^\dagger}$. The TM of the combined MMF-inverter system is given by ${\VecT_{\text{fib-inv}} = \VecT_{\text{inv}}\VecT_{\text{fib}}= \VecP\VecP^\dagger}$, which now represents only the spatial filtering applied to light fields transmitted through a fibre (due to the fibre's limited $\text{NA}$), with spatial scrambling corrected.

The PIMs spatially overlap with one another, so separating them to impart the required mode-dependent phase delay onto each PIM is the key challenge we face in the design of an optical inverter~\cite{butaite2022build}. The technique of multi-plane light conversion is emerging as a front-runner to efficiently manipulate tens of arbitrarily shaped spatial light modes simultaneously~\cite{morizur2010programmable,wang2018dynamic,fontaine2019laguerre,brandt2020high,fickler2020full,mounaix2020time,lib2021reconfigurable,kupianskyi2023high,korichi2023high}. Here we aim to design a multi-plane light converter (MPLC) that has a relatively low number of planes, rendering the design practical to build: we unscramble the $N=42$ spatial channels using a cascade of $M=5$ phase planes.

The phase profiles of the layered MPLC structure can be efficiently inverse-designed using methods analogous to the back-propagation algorithms used to train layered electronic artificial neural networks~\cite{hashimoto2005optical,fontaine2019laguerre,barre2022inverse}. Transforming many spatial modes with only a few phase planes typically means that not all of the input light can be fully controlled. This situation is dependent upon the desired transform asked of the MPLC, but typically occurs if the number of planes $M\lesssim 2N$. To overcome this problem, here we employ a bespoke MPLC design algorithm that allows a high degree of tunability in the low-plane number regime. We recently introduced this algorithm for generalised mode sorting~\cite{kupianskyi2023high}, and here we apply it to optical inverter design. Our iterative algorithm relies on gradient ascent with an objective function that enables the trade-off between fidelity and efficiency to be adjusted on a mode-by-mode basis -- see Methods and ref.~\cite{kupianskyi2023high} for a detailed description. SI \S 3 shows a comparison of our gradient ascent algorithm to conventional approaches -- in particular, the well-known `wavefront matching method' (WMM)~\cite{hashimoto2005optical,fontaine2019laguerre}. These simulations demonstrate that our gradient ascent algorithm gives access to substantially higher fidelity inverter designs in this scenario.

Figure~\ref{Fig:svd_inv}(a-b) shows a simulation of the performance of an optical inverter designed using gradient ascent. As can be seen in Fig.~\ref{Fig:svd_inv}(a), the target output PIMs are projected into a disk in the output plane, and uncontrolled light is directed around the edge of this disk where it can be discarded or blocked. When the optical inverter is coupled to an MMF, this output disk becomes an image of the field illuminating the input facet of the fibre. The optical inverter will unscramble spatially coherent laser light, or spatially incoherent (or partially coherent) light, within the spectral bandwidth of the combined MMF-inverter system. Figure~\ref{Fig:svd_inv}(b) shows simulated examples of spatially incoherent images transmitted through the combined MMF-inverter system with high fidelity. The resolution of these images are diffraction limited, governed by the NA of the MMF.

The incoherent imaging capabilities of this system can be understood by considering that light emanating from a diffraction limited point on the input facet of the MMF is re-imaged to a corresponding point at the output of the inverter. Therefore, the operation of the system does not rely on interference between light from neighboring points, meaning there is no requirement for spatial coherence of the input optical field. The bandwidth, $\Delta\lambda$, is thus limited by the spectral dispersion of the combined system, which for a step-index fibre, is typically constrained by the fibre itself ($\Delta\lambda_{\mathrm{fib}}$). For imaging at the output facet of a step-index fibre, $\Delta\lambda_{\mathrm{fib}}\sim2n_c\lambda^2/(L\mathrm{NA}^2)$. Although relatively narrow, this spectral bandwidth substantially increases if the image plane is moved away from the end of the fibre, as discussed in ref.~\cite{butaite2022build}. Therefore, far-field imaging through MMFs~\cite{leite2021observing,stellinga2021time} may be achieved over a broad spectral bandwidth. SI \S 6 shows simulations of the spectral bandwidth of the optical inverters designed in this work.\\

\noindent{\bf Singular value decomposition-based inverter design}:
Next, we study the design of an optical inverter matched to a real step-index MMF of 1\,m in length, nominally with the same core diameter and $\text{NA}$ as simulated above (${d=25\pm3\,\mu\text{m}}$, ${\text{NA}=0.1}$). We first experimentally measure the TM of this fibre, $\VecT_{\text{f-exp}}$, at a wavelength of 633\,nm, in a single circular input and output polarisation state~\cite{popoff2010measuring,vcivzmar2012exploiting}. A digital micro-mirror device (DMD), placed in the Fourier plane of the input facet and acting as a programmable diffraction grating, is used to shape the laser light projected into the fibre~\cite{lee1979binary,mitchell2016high}. The TM is measured by scanning a focused spot over a hexagonal grid of points across the input facet, and holographically recording the fields emanating from the output facet. See Methods for experimental details, and SI \S 1 for a schematic of the full optical set-up.

To represent a realistic use-case, the fibre is held in place in a curved configuration (as shown in SI \S 2) and so the TM features non-negligible levels of spatial mode and polarisation coupling. Our prototype optical inverter is designed to operate on a single polarisation state, and so we filter out one circular polarisation of light exiting the fibre, rendering the measured TM non-unitary. We note that a polarisation-resolved TM of a short fibre is typically close to unitary~\cite{ploschner2015seeing}, and our approaches could be naturally extended to vectorial optical inverters capable of operating on both polarisation states simultaneously~\cite{fontaine2017programmable}. 

Rather than the eigenvalue decomposition applied above, it is now more appropriate to design the optical inverter by considering the singular value decomposition (SVD) of the fibre TM: ${\VecT_{\text{f-exp}} = \VecU\boldsymbol{\Sigma}\VecV^\dagger}$, where unitary matrices $\VecU$ and $\VecV$ contain the left-hand and right-hand singular vectors, respectively, along their columns, and diagonal matrix $\boldsymbol{\Sigma}$ contains the singular values along its diagonal. Figure~\ref{Fig:svd_inv}(e) shows a plot of the first 100 singular values in descending order. We see the distribution of singular values is dominated by $\sim30$ large values, corresponding to speckle field profiles that approximate linear combinations of ideal PIMS. These speckle patterns transmit the majority of the power through the MMF. This number of high singular values agrees well with the theoretical mode capacity calculated from the fibre geometry when also factoring in the manufacturing tolerance on core radius. We observe a long tail of lower singular values, a phenomenon which we interpret as due to core-cladding modes that are weakly excited by light scattering out of the core.

In this scenario, the inverse transform that our optical inverter must apply can be approximated by the pseudo-inverse of $\VecT_{\text{f-exp}}$, i.e.\ ${\VecT_{\text{f-exp}}^{-1}\sim\VecV\VecS^{-1}\VecU^\dagger}$. Here we regularise the inverse by setting all but the largest $N=30$ singular values to zero. Therefore, we design an MPLC to simultaneously pair-wise map the $N$ spatial modes represented by the left-hand singular vectors (held on columns of $\VecU$) with the largest singular values, to the corresponding $N$ spatial modes defined by the right-hand singular vectors (held on columns of $\VecV$). As before, when reformatted to a 2D array, these right-hand singular vectors fit into a disk representing an image of the input facet of the fibre formed at the output of the inverter. Examples of these left and right singular vectors are shown in Fig.~\ref{Fig:svd_inv}(c) and SI \S 5, along with simulated outputs from the SVD-based inverter.

\begin{figure*}[t]
   \includegraphics[width=1.0\textwidth]{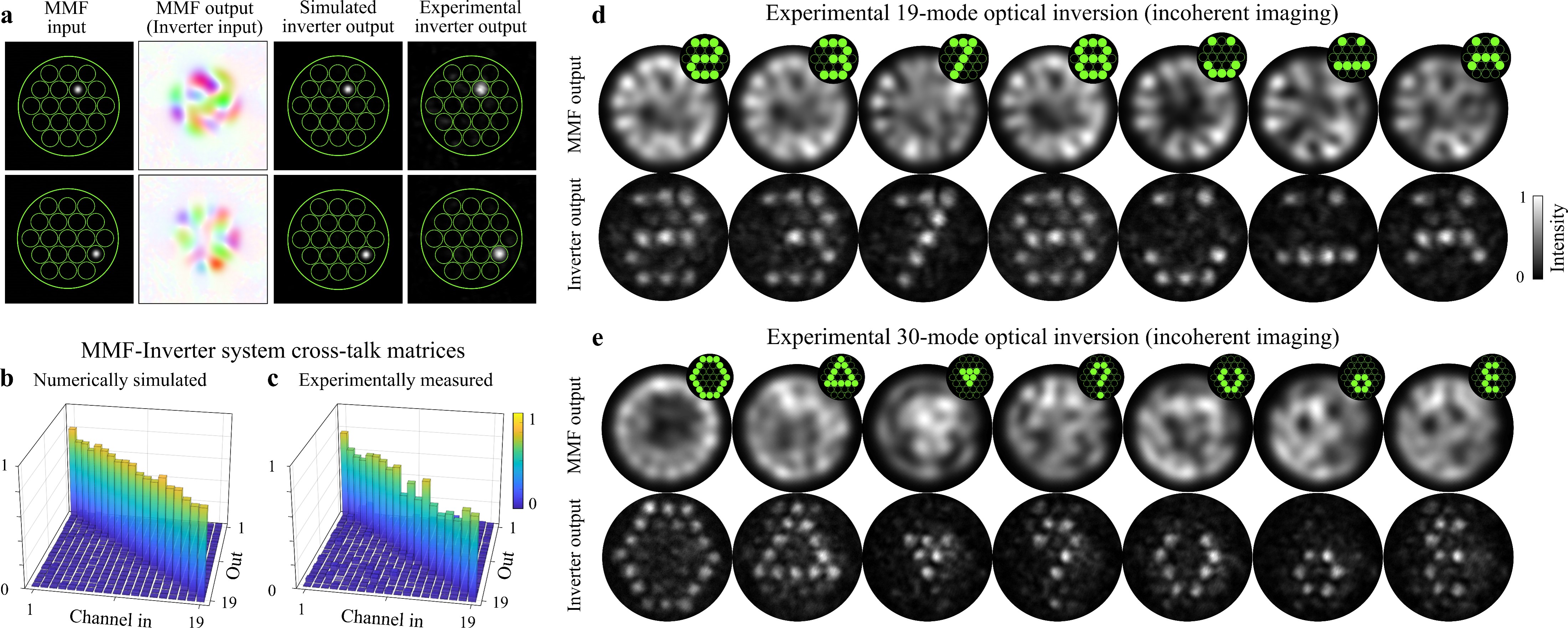}
   \caption{{\bf Experimental all-optical MMF inversion}. (a) Operation of a 19-channel inverter, formed from 5 phase planes with a resolution 200$\times$200 pixels (examples of optimised phase profiles shown in Fig.~\ref{Fig:concept}(b)). Left-hand column: Two examples of focused spots illuminating the input fibre facet, intensity shown. The locations of the 19 input channels within the core are shown with green circles. Second column from left: the resulting speckle patterns generated at the MMF output, experimentally measured optical fields shown. Third column: simulated inverter output, intensity shown. Right-hand column: experimentally measured inverter output, intensity shown. (b) Simulated channel cross-talk matrix for the 19-channel inverter design. Mean cross-talk is $c\sim 24\,\%$, and mean efficiency is $e\sim27\,\%$ (see Methods for definitions). (c) Experimentally measured channel cross-talk matrix for the 19-channel inverter design. Mean cross-talk is $c\sim 34\,\%$. (d) Incoherent images transmitted through the prototype 19-channel fibre-inverter system. In each case, a subset of the input channels are excited as shown in the insets (top right of each panel). Upper row: scrambled intensity pattern at output fibre facet. Lower row: intensity at inverter output. (e) Incoherent images transmitted through a 30-channel fibre-inverter system.}
   \label{Fig:exp_inv}
\end{figure*}

As the TM of the MMF is non-unitary, the singular vectors do not have a uniform magnitude. In this situation an ideal inverter would somehow boost the power in the singular vectors with lower singular values to compensate for this effect -- a feature that at first sight appears incompatible with a passive linear optical system. However, given a low-plane MPLC is inherently lossy, and we have independent control over the transform efficiency of each singular vector, our algorithm can, within limits, be used to re-balance these mode dependent losses by boosting the transform efficiency of certain modes -- see Methods.

Figure~\ref{Fig:svd_inv}(d) shows the simulated performance of this SVD-based optical inverter design, when matched to the experimentally measured fibre TM. In this case the fidelity of imaging is slightly reduced -- in particular on the left-hand-side of the field-of-view (FoV). This is due to the non-unitary nature of the single-polarisation channel TM. In particular, some of the light transmitted from the left-hand-side of the input facet is transformed into the polarisation channel, or into singular vectors that are filtered out. Characterising and inverting both polarisation channels simultaneously could overcome this loss of information~\cite{fontaine2017programmable}.\\ 

\noindent{\bf Optical inversion via speckle mode sorting}:
So far we have designed optical inverters capable of reconstructing images of arbitrary continuous fields incident anywhere on the input fibre facet. However, in the more realistic non-unitary case this gives us no direct control over the spatial variation in reconstruction fidelity. In our final strategy, we investigate an inversion method that allows us to arbitrarily specify the regions of the fibre facet that we wish to reconstruct with high fidelity.

Our measured fibre TM $\VecT_{\text{f-exp}}$ maps a set of input focused spots to the corresponding speckle patterns emerging from the other end of the fibre. We now design an MPLC to perform the inverse mapping, and directly transform these speckle patterns back into focused spots. In this design protocol, we can arbitrarily specify the number and position of diffraction limited spots at the input facet which are unscrambled and re-imaged to the output of the inverter. For example, we are able to lower the number of channels to fewer than the maximum supported by the fibre, which allows a smaller set of channels to be unscrambled with higher fidelity.

If the input channels (i.e.\ focused spots) are not spatially overlapping, then the performance of this `speckle mode sorter' can be quantified by a cross-talk matrix, which depicts the fraction of light input into channel index $i$ at the MMF input facet that appears in output channel index $j$ after the inverter (see Methods). Figure~\ref{Fig:exp_inv}(b) shows the numerically simulated cross-talk matrix of a speckle sorter-based inverter designed to operate on 19 spatially separated channels using 5 phase planes. Here we limit the design to an experimentally achievable resolution to compare with the experiments detailed below. In this case we design the speckle mode sorter by optimising the correlation between the target and actual output modes, which is equivalent to the WMM. SI Movie 1 shows a simulation of the spatial transformation undergone by input speckle fields as they propagate through the phase planes of a speckle mode-sorter inverter design. We note that our tunable inverse design algorithm can be used to substantially suppress cross-talk in future high-resolution speckle mode sorter designs~\cite{kupianskyi2023high}.  \\

\noindent{\large{\bf Experimental realisation}}\\
We now experimentally implement a prototype optical inverter matched to the 1\,m long MMF characterised earlier. Figure~\ref{Fig:concept} shows a simplified schematic of our experiment, and SI \S 1 shows the full experimental setup. Our inverter is constructed from an MPLC with 5 reflections from a liquid crystal SLM. The phase planes are designed using knowledge of the experimentally measured fibre TM ${\VecT_{\text{f-exp}}}$. The output fibre facet is magnified and imaged onto the first phase plane of the MPLC. When processing tens of spatial light modes, the complexity of the required phase profiles means that pixel-perfect alignment of each phase plane is critical. This is demanding to achieve given the numerous alignment degrees-of-freedom (as discussed in detail in ref.~\cite{kupianskyi2023high}). To mitigate alignment difficulties, we create a range of MPLC designs incorporating the expected span of residual experimental errors in the scaling and defocus of the input fields, and distance between planes, and implement an auto-alignment protocol based on a genetic algorithm that simultaneously optimizes the choice of phase-plane set and the lateral position of each phase plane on the SLM. Once the fibre is secured in position on an optical table, and the MPLC is aligned, we find the system is stable for days at a time.

Experimentally, the resolution of the SLM limits us to relatively low resolution MPLC designs, which best suit the speckle-sorter based optical inverter, since the SVD-based design requires much higher resolution phase patterns~\cite{kupianskyi2023high}. We first test the performance of the inverter to unscramble 19 channels evenly distributed over a hexagonal grid across the input facet of the fibre. Figure~\ref{Fig:exp_inv}(a) shows two examples of input spots transformed into speckle patterns via propagation through the MMF, and then back into focused spots by the optical inverter. Our experimentally realised device compares well with simulations. Figures~\ref{Fig:exp_inv}(b-c) show a comparison of the simulated and experimentally measured cross-talk matrices in this case. The slightly higher system cross-talk levels observed experimentally are mainly due to the effects of coupling between adjacent SLM pixels (i.e.\ the phase setting of one pixel affects the phase of neighbouring pixels), which may be mitigated by using higher-resolution or multiple SLMs to display the required phase profiles with greater accuracy. SI Movies 2 and 3 show the inverter outputs as all input channels are sequentially excited, for a 19-mode and 30-mode inverter.\\

\begin{figure*}[t]
   \includegraphics[width=0.85\textwidth]{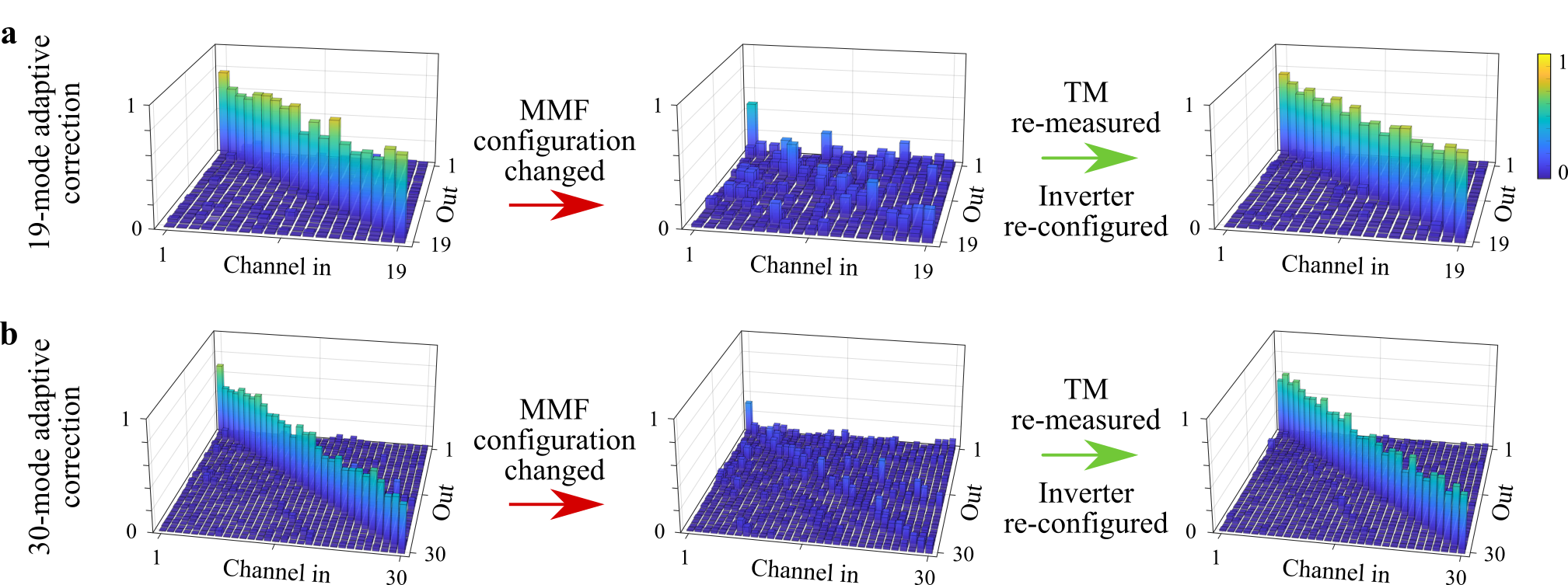}
   \caption{{\bf Adaptive optical inversion.} (a) Cross-talk matrices of a 19-channel MMF-inverter system. Left-hand column: initial aligned state. Middle column: after fibre perturbation, with inverter still set to initial state. Right column: inverter adapted to new fibre state. (b) Equivalent plots to (a), but for a 30-channel inverter. The average cross-talk values, from left to right, for (a) are $c \sim 34\,\%,~86\,\%$ and $33\,\%$ respectively, and for (b) $c \sim 49\,\%,~91\,\%$ and $48\%$.}
   \label{Fig:exp_inv_bend}
\end{figure*}

\noindent{\bf All-optical image transmission through MMFs}: Figure~\ref{Fig:exp_inv}(d-e) demonstrates all-optical incoherent image transmission through the MMF-inverter system. We design and implement optical inverters supporting both 19 channels (Fig.~\ref{Fig:exp_inv}(d)) and 30 channels (Fig.~\ref{Fig:exp_inv}(e)). In these experiments, we mimic the effect of incoherent light transmitted through the system by illuminating each excited channel sequentially with the DMD, and time-averaging the intensity images at the output of the inverter -- this captures the level of cross-talk expected for the transmission of incoherent light within the spectral bandwidth of the system. We transmit a variety of simple pixellated binary images of numbers, letters, smiley faces and symbols, and observe that all images are recognisable without the need for any additional processing at the output of the inverter. The contrast of the transmitted images reduces as their sparsity decreases (i.e.\ as more channels are simultaneously excited) -- as is expected from the additive effects of incoherent channel cross-talk. Imaging contrast also reduces when the number of channels is increased from 19 to 30, due to the additional load placed on the MPLC-based speckle sorter to simultaneously sort more modes in 5 phase planes.\\

\noindent{\bf Adaptive optical inversion}: As with all approaches that rely on knowledge of the TM of a scattering medium, our inversion strategy will fail if the fibre configuration changes enough to significantly modify the TM. The transmission properties of optical fibres are notoriously sensitive to such perturbations. Therefore, if changes are anticipated, the fibre TM should be regularly re-estimated (which can be achieved with only a few probe measurements~\cite{li2021compressively}) and the MPLC patterns updated when needed. We explore this scenario in Fig.~\ref{Fig:exp_inv_bend}. Starting from an aligned MMF-inverter system (Fig.~\ref{Fig:exp_inv_bend}, left-hand column), we deliberately change the fibre configuration by re-routing one of its bends (see SI \S 2 for details). This almost entirely disrupts the unscrambling operation, as seen by the disappearance of the prominent diagonal in the cross-talk matrices shown in Fig.~\ref{Fig:exp_inv_bend}, middle column. Performance is restored by remeasuring the TM of the MMF, and re-configuring the phase-planes of the optical inverter to match the new TM, as shown in Fig.~\ref{Fig:exp_inv_bend}, right column.

We note that in micro-endoscopy applications, the input end of the fibre may not be optically accessible -- thus complicating continuous TM monitoring of flexible MMFs. Nonetheless, there are a variety of methods under development to achieve single-ended TM measurement in this scenario~\cite{gu2015design,gordon2019characterizing,li2021memory,wen2023single}. Even once a new TM is known, in future adaptive applications it will be important to minimise the computational complexity associated with re-designing an entire optical inverter on-the-fly whenever the fibre state changes. To address this issue, we showed in ref.~\cite{butaite2022build} that modulating only a single, carefully chosen phase plane can correct for a wide range of fibre states. Furthermore, selecting from a pre-designed library of inverters, matched to the range of expected fibre TMs, would bypass the need for any synchronous inverter design~\cite{wen2023single}, offering a viable route towards adaptive inversion operating at liquid crystal SLM update rates in the future.\\

\noindent{\large{\bf Discussion}}\\
We now consider our work in the context of other emerging techniques and give an outlook towards future directions. An MPLC is equivalent to the more recently coined {\it diffractive artificial neural network} (DNN)~\cite{lin2018all} -- the two concepts share the same physical structure, and are inverse designed using analogous approaches. DNNs have recently begun to be applied to computational imaging operations~\cite{luo2022computational,mengu2022all}. Viewed from the perspective of artificial neural networks, an ideal MMF optical inverter is capable of reconstructing any transmittable image after being `trained' (i.e.\ inverse-designed) on a minimal yet complete set of fibre responses. This minimises the training time, and prevents any undesired bias towards particular classes of image being frozen into the final network design.

More crucially, the all-optical nature of our physically-realised network conserves the phase and coherence information carried by optical fields flowing through it. This responsiveness to the optical phase renders the inversion problem linear and well-posed, offering advantages over conventional neural networks tasked with this class of light unscrambling problem. For example, electronic neural networks have been applied to unscramble coherent images transmitted through MMFs~\cite{borhani2018learning,caramazza2019transmission}. When trained on intensity-only images cast by spatially coherent light, these methods are faced with a non-linear mapping problem that is highly ill-posed, as many different inputs can result in the same intensity profile at the output of the fiber, differing only by their phase information~\cite{gigan2022roadmap}. That said, conventional neural networks currently offer additional flexibility in terms of accommodating perturbations to fibre state~\cite{resisi2021image,abdulaziz2022robust}.

Single-shot incoherent computational imaging directly through MMFs has also been previously explored. This technique relies on measurement of the `intensity TM', $\VecM$, which links incoherent optical fields at either end of a fibre~\cite{kim2014ultra,butaite2022build}. Measurement of the intensity pattern  of a spatially incoherent field at one end of the fibre, $\Veca$, then permits the intensity pattern at the other end, $\Vecb$, to be computed by solving the inverse problem ${\VecM\Vecb = \Veca}$ for $\Vecb$. However, in this approach, the spatial information is encoded in low contrast speckle patterns such as those shown in Fig.~\ref{Fig:exp_inv}(c,d) (top rows). This means that matrix $\VecM$ is poorly conditioned, and the solution to the inverse problem is very sensitive to small changes in $\Veca$. For example, we see that in Fig~\ref{Fig:exp_inv}(d) (top row), the speckle patterns emanating from the fibre look very similar when transmitting images of `2', `3' and `8' -- making accurate computational reconstruction of these images sensitive to noise. Therefore, this direct inversion technique is hampered by very low signal-to-noise ratios, and typically works best when imaging sparse scenes~\cite{guo2022scan}.

An ideal inverter physically solves this incoherent imaging inverse problem without the need for any further computation. However, even an imperfect inverter, possessing non-negligible levels of channel cross-talk (such as our 30-channel prototype), has advantages here: it represents a {\it physical pre-conditioning element}, thus reducing the condition number (i.e.\ the ratio of the largest to the smallest singular value) of the intensity TM describing the optical system. This improves the stability of final computational image reconstruction. For example, in our experiments the inclusion of the inverter lowered the condition number $\kappa$ of matrix $\VecM$ from $\kappa=14.1$ (MMF alone) to $\kappa=5.5$ (MMF-inverter) in the 30-channel case, and from $\kappa=8.8$ to $\kappa=3.1$ in the 19-channel case (see Methods). We anticipate the benefits of physically pre-conditioning in this way will grow with the dimensionality of the system.

How scalable is our concept in terms of the length and mode capacity of the fibre? The main challenge presented by longer fibres is the stability of their TM, which becomes increasingly susceptible to external perturbations (e.g.\ changes in temperature, or vibrations). Therefore long fibres would necessitate low latency adaptive optical inversion. Regarding mode capacity: in this work we experimentally unscramble up to 30 modes using 5 phase planes, i.e.\ a mode-to-plane ratio of 6, with an efficiency of $\sim20\%$ (ignoring SLM losses). SI \S 7 provides a table of performance metrics for all of the inverters designed in this work. In the general case of arbitrary unitary transformations, to unscramble light with high fidelity and efficiency, the number of planes scales linearly with the number of modes~\cite{pastor2021arbitrary}. However, this scaling can be improved by sacrificing transform efficiency while preserving fidelity~\cite{kupianskyi2023high}. Interestingly, certain transformations can be efficiently achieved with a far more favorable mode-to-plane scaling~\cite{berkhout2010efficient,fontaine2019laguerre}. We recently showed that the transformation required to sort MMF PIMs falls into this category~\cite{butaite2022build}. In ref.~\cite{butaite2022build} we constrained the inverter design to take advantage of the efficient PIM sorting transform, and established that, in theory, up to 400 modes could be unscrambled in 29 planes with an efficiency of $\sim50\%$. This improves the mode-to-plane ratio to $\sim$14. Experimentally realising such a high-dimensional system is a promising avenue for future work.\\

\noindent{\large{\bf Conclusions}}\\
In summary, we have demonstrated a passive optical system capable of reversing the strongly spatially-variant aberrations introduced by a MMF -- unscrambling tens of optical modes simultaneously. Our approach can be understood as a form of all-optical MIMO equilisation~\cite{tanomura2020monolithic}. Proposals to optically unscramble light propagation through MMFs were first suggested in the 1970s~\cite{gover1976direct}. However, it is only in the last few years that our understanding of the inverse design of multi-modal photonic systems has matured to a level that renders such concepts experimentally feasible. In addition to the step-index MMFs studied here, these approaches apply to graded-index fibres, fibre bundles and photonic lanterns~\cite{badt2022real,choudhury2020computational}. Our work targets micro-endoscopic imaging applications, however these concepts apply more broadly to general scattering media~\cite{kang2023tracing}, and also have potential applications in the fields of classical and quantum optical communications and photonic computing.

\section{Methods}
\noindent{\bf Optical inverter inverse design algorithm}\\
Our aim is to find the phase delay imparted by each pixel on each plane of the optical inverter, such that the resulting optical transformation is close to the desired inverter transmission matrix $\VecT_{\text{inv}}$. For each design strategy, we specify different target sets of input and output modes -- essentially selecting the preferred basis in which the inverter design will be conducted:
\begin{itemize}
    \item For the ideal Eigenmode-based inverter design, the input mode set is the PIMs -- see, for example, refs~\cite{ploschner2015seeing} or~\cite{li2021memory} for details of how these modes are defined, and SI \S 4 for visualisations. The output set is also the PIMs, with the global phase shift of the $n^{\text{th}}$ mode given by ${\theta_n = -\beta_nL}$, to compensate for the phase delay picked up during propagation through a fibre of length $L$.
\item For the SVD-based inverter, we express the measured TM as ${\VecT_{\text{f-exp}} = \VecU\boldsymbol{\Sigma}\VecV^\dagger}$, as explained in the main text. The inverter input modes are the left-hand singular vectors held on the columns of $\VecU$, and the output modes are the right-hand singular vectors held on the columns of $\VecV$ -- appropriately truncated.
\item For the speckle mode sorter, the input modes are user-selected columns of $\VecT_{\text{f-exp}}$ -- i.e.\ experimentally measured speckle patterns emanating from the fibre during the measurement of the TM. The output modes are the corresponding focused points that were focused onto the input of the fibre to probe the fibre TM. 
\end{itemize}
During the design process, we aim to optimise a metric that quantifies the performance of the MPLC to transform all input modes to their corresponding output modes -- taking into account both the transform fidelity and efficiency of each mode pair. To achieve this, we use the efficient MPLC design method we recently introduced in ref.~\cite{kupianskyi2023high}, and choose the optimisation objective function as follows
\begin{gather}\label{Eqn:TotalObjective}
F_\text{T} = \sum_{n=1}^N \alpha_nF_n\,,\\
F_n = \underbrace{{\rm Re}\left[\Vecv^{\dagger}_n\cdot\boldsymbol{\rho}_n\right]}_\text{Fidelity}+\underbrace{\gamma_n\left(\Vecv^{\dagger}_n\cdot\Vecv^{\text{bk}}_n\right)}_\text{Efficiency}\,.\end{gather}
Here $F_\text{T}$ is the real-valued number we aim to maximise during optimisation. $N$ is the total number of mode-pairs to transform. The real positive number $\alpha_n$ weights the relative importance of transforming mode pair $n$. 

$F_n$ is a real-valued number quantifying the performance of the transformation for input-output mode-pair $n$. The interplay between the two contributions to $F_n$ can be understood as follows: The first term on the right-hand-side (RHS) of Eqn.~2 is designed to maximise the correlation between the target output mode $\boldsymbol{\rho}_n$, and the actual output mode $\Vecv_n$. Taking the real part of the overlap constrains the global phase of the output to match the global phase of the target (alternatively, taking the absolute square of this term leaves the relative global phase of the output modes unconstrained~\cite{barre2022inverse,kupianskyi2023high}).

The second term on the RHS of Eqn.~2, weighted by positive scalar $\gamma_n$, enables the efficiency of the transformation to be tuned. A non-zero value of $\gamma_n$ allows some of the transmitted light to be deliberately shepherded into a designated background region outside the image of the input fibre facet. This lowers the overall efficiency, but enables optical inverter designs to be found that yield a higher output fidelity within the image of the fibre facet itself. The background region is defined using vector $\Vech^{\text{bk}}$: the elements of $\Vech^{\text{bk}}$ are set to zero inside the image of the fibre core where the unscrambled field will appear, and 1 in the surrounding area, which is designated as the background zone. We incorporate this information into Eqn.~2 via ${\Vecv^{\text{bk}}_n = \Vecv_n\odot\Vech^{\text{bk}}}$, where the operation $\odot$ signifies the element-wise Hadamard product. Therefore ${\Vecv^{\dagger}_n\cdot\Vecv^{\text{bk}}_n}$ represents the intensity of light directed to the background zone (and is thus always real and positive). Adding this efficiency term in Eqn.~2 acts to increase $F_n$ when more light is scattered to the background. We note that the spatial structure of $\Vecv_n$ in the background zone is free to evolve throughout the iterative design process, so this approach does not enforce any predetermined structure on the light scattered there.

Within Eqn.~2, when the efficiency term increases, the fidelity term simultaneously decreases, since here the calculation of fidelity involves the overlap integral between the target and actual mode across the {\it entire} output plane. Thus, this formulation allows the relative importance of fidelity and efficiency to be tuned on a mode-by-mode basis by the adjusting the values of $\gamma_n$. In addition, the relative importance of each input-output mode-pair can be adjusted by tuning $\alpha_n$. The combination of these two adjustable controls enables the efficiency of certain mode-pair transformations to be boosted -- thus correcting, to some extent, for the non-uniform singular value distribution of the inverse TM that should be implemented in the SVD-based optical inverter design.

For the eigenmode-based and SVD-based inverter designs we use non-zero values of $\alpha_n$ and $\gamma_n$, that are initialised as $\alpha_n = 1$ and $\gamma_n = 2$, while $\gamma_n$ is adjusted throughout the design process to compensate for the variation in transformation weights governed by the diagonal elements of $\boldsymbol{\Sigma}$. For the speckle mode sorter designs, we use $\alpha_n = 1$ and $\gamma_n = 0$. In this case, the algorithm optimises the same objective as the well-known wavefront matching method~\cite{hashimoto2005optical,fontaine2019laguerre}. If higher resolution phase masks are possible to implement, the cross-talk between the speckle sorter channels can be substantially suppressed by adding an additional term to the objective function, as shown in ref.~\cite{kupianskyi2023high}.

Equation 2 is readily differentiable with respect to the phase profile on a particular plane, meaning that the objective function can be efficiently maximised using gradient-based adjoint methods~\cite{kupianskyi2023high}.
\\

\noindent{\bf Optical inverter performance metrics}\\
The performance of a particular optical inverter design can be quantified in several ways, described in detail in the supplementary information of ref.~\cite{kupianskyi2023high}, and summarised below:

\noindent {\it Efficiency}: The efficiency $e_n$ of the transformation of mode $n$ is given by fraction of input power transmitted into the target zone at the inverter output. For the case of the Eigenmode and SVD-based inverters, this output zone is the disk representing the image of the fibre input facet. For the speckle mode sorter inverter design, the output zone is a small disk located where the output Gaussian should be focused to. The mean efficiency $e$ is given by the average of $e_n$ over all $N$ modes.

\noindent {\it Fidelity}: The fidelity $f_n$ of the transformation of mode $n$ is given by the absolute square of the normalised overlap integral between the target spatial mode, and the actual spatial mode that is transmitted into the target output zone (i.e.\ setting any part of the actual field outside the target zone to zero before normalisation takes place). The mean fidelity $f$ is given by the average of $f_n$ over all $N$ modes.

\noindent {\it Channel cross-talk}: For the speckle mode sorter-based inverter, if the output channels do not spatially overlap, we can quantify the level of channel cross-talk. This information is stored in a matrix $\Vecc$, where element $c_{i,j}$ is given by the amount of power appearing in output channel $j$, when input channel $i$ is excited, divided by the total power appearing in all output channels. In the ideal case of no channel cross-talk, $c_{i,i} = 1$ for all $i$, and $c_{i,j} = 0$ for all $i\neq j$, meaning that $\Vecc$ is equal to the identity matrix. The mean cross-talk is given by $c$, which is equal to one minus the mean value of the diagonal elements of $\Vecc$ (represented as a percentage in the main text) i.e.\ $c=0\,\%$ in the ideal case.\\

\noindent{\bf Fibre transmission matrix characterisation}\\
 Our experimental set-up for measurement of the TM of the MMF is similar to that described in ref.~\cite{li2021compressively}. The MMF used in our experiments is a 1\,m long step-index MMF with a nominal core diameter of $d=25\,\mu$m, and numerical aperture $\text{NA}=0.1$ (Thorlabs FG025LJA). A 633\,nm laser beam, generated by a 1\,mW HeNe laser, (Thorlabs HNLS008L-EC), is expanded to fill a DMD (Vialux-7001), which is used to shape the light incident on the input end of the fibre~\cite{lee1979binary,mitchell2016high}. The input facet of the fibre is placed in the Fourier plane of the DMD. The incident light is circularly polarised. Light is focused into, and collected from, the MMF using a pair of 10$\times$ objective lenses. The output facet of the fibre is imaged onto a camera that is electronically synchronised with the DMD. A combination of a quarter wave-plate and a polariser filter out one component of circular polarisation. A coherent reference beam (plane wave) is also incident onto the camera, enabling measurement of the optical field via digital holography.
 
 We measure the TM $\VecT_{\text{f-exp}}$ of this fibre by scanning a focused spot over a hexagonal grid across the input facet. Each of these inputs results in a unique speckle pattern emerging from the output fibre facet. Each output field is vectorised and the $p^{\text{th}}$ output forms column $p$ to the TM $\VecT_{\text{f-exp}}$. We implement phase-drift correction to compensate for any phase drift between the signal and reference arms of the interferometer. This is achieved by interlacing the probe measurements with a standard measurement. The global phase drift of this standard output mode is tracked throughout the TM measurement, and the phase drift function subtracted from the phase of the final TM measurement. SI \S 1 shows a detailed schematic of the optical set-up.\\

 \noindent{\bf Experimental implementation of optical inverter}\\
Once designed, we implement the optical inverter using five reflections from a liquid crystal SLM (Hamamatsu X13138-01), of total resolution 1280$\times$1024 pixels and a pixel pitch of $12.5~\mu m$. The SLM is placed opposite a mirror, positioned a distance of 31\,mm away from the SLM chip, giving a distance between the phase planes of 62\,mm. Each phase plane is first optimised using a 400$\times$400 pixel simulation. The active area of each plane, where the light from each mode mainly stays, is a central 200$\times$200 pixel region. In order to fit the 5 phase planes adjacently on a single SLM, we crop each pre-designed phase plane to a 200$\times$400 pixel area, which are displayed next to one another on the SLM. Initial experimental alignment of this optical system is challenging, as pixel-perfect lateral positioning of the phase planes is required, leading to a large number of degrees of freedom to align simultaneously. We achieved this by implementing an automated genetic algorithm to search for the correct phase plane display positions on the SLM. See ref.~\cite{kupianskyi2023high} for a detailed description. The output of the inverter is recorded with a camera (Basler ace acA640-300gm), positioned in the Fourier plane of the final phase plane. The operation of this camera is electronically synchronised with the DMD.

We note that the SLM used for this work did not have its reflection efficiency optimised specifically for the operating wavelength of 633\,nm. Therefore the efficiency of each reflection was relatively low ($\sim$50\,\%), thus rendering the efficiency of the optical inverter artificially low in our prototype device. This issue could be improved by using an SLM with a dielectric back-plane optimised for the operational wavelength.\\

\noindent{\bf Condition number of the intensity transmission matrix}\\
In the main text we analyse the condition number $\kappa$ of the intensity TM $\VecM$ of just the MMF ($\VecM_{\text{MMF}}$), and of the MMF-inverter system ($\VecM_{\text{MMF-inv}}$). In general, the intensity TM of a scattering system is given by $\VecM = |\VecT|^2$, where $\VecT$ is the field TM when represented in real-space input and output bases, and here we take the element-by-element (i.e.\ Hadamard) square. Therefore, the $n^{\text{th}}$ column of $\VecM_{\text{MMF}}$ is given by the vectorised intensity speckle pattern that appears at the output facet of the MMF, when the MMF is excited with the $n^{\text{th}}$ input focused spot on the input facet. Similarly, the $n^{\text{th}}$ column of $\VecM_{\text{MMF-inv}}$ is given by the vectorised intensity pattern that appears at the output of the inverter when the MMF is excited with the $n^{\text{th}}$ input focused spot incident onto the input facet. The condition number $\kappa$ of these matrices is then calculated by finding the singular value decomposition of $\VecM$, and calculating the ratio of the largest singular value divided by the smallest singular value.

\section{Acknowledgements}
We thank Unė Būtaitė, Tomáš Čižmár and Joel Carpenter for useful discussions. DBP acknowledges financial support from the European Research Council (Grant no.\ 804626), and the
Royal Academy of Engineering.

\section{Contributions}
DBP conceived the idea for the project and supervised the work. HK performed all simulations, experimental work and data analysis. SARH derived the gradient ascent optimisation method for Eigenmode and SVD-based optical inverter designs. DBP and HK wrote the paper, with editorial input from SARH.

\bibliography{expOpticalInverterRefs}

\begin{thebibliography}{71}%
\makeatletter
\providecommand \@ifxundefined [1]{%
 \@ifx{#1\undefined}
}%
\providecommand \@ifnum [1]{%
 \ifnum #1\expandafter \@firstoftwo
 \else \expandafter \@secondoftwo
 \fi
}%
\providecommand \@ifx [1]{%
 \ifx #1\expandafter \@firstoftwo
 \else \expandafter \@secondoftwo
 \fi
}%
\providecommand \natexlab [1]{#1}%
\providecommand \enquote  [1]{``#1''}%
\providecommand \bibnamefont  [1]{#1}%
\providecommand \bibfnamefont [1]{#1}%
\providecommand \citenamefont [1]{#1}%
\providecommand \href@noop [0]{\@secondoftwo}%
\providecommand \href [0]{\begingroup \@sanitize@url \@href}%
\providecommand \@href[1]{\@@startlink{#1}\@@href}%
\providecommand \@@href[1]{\endgroup#1\@@endlink}%
\providecommand \@sanitize@url [0]{\catcode `\\12\catcode `\$12\catcode
  `\&12\catcode `\#12\catcode `\^12\catcode `\_12\catcode `\%12\relax}%
\providecommand \@@startlink[1]{}%
\providecommand \@@endlink[0]{}%
\providecommand \url  [0]{\begingroup\@sanitize@url \@url }%
\providecommand \@url [1]{\endgroup\@href {#1}{\urlprefix }}%
\providecommand \urlprefix  [0]{URL }%
\providecommand \Eprint [0]{\href }%
\providecommand \doibase [0]{http://dx.doi.org/}%
\providecommand \selectlanguage [0]{\@gobble}%
\providecommand \bibinfo  [0]{\@secondoftwo}%
\providecommand \bibfield  [0]{\@secondoftwo}%
\providecommand \translation [1]{[#1]}%
\providecommand \BibitemOpen [0]{}%
\providecommand \bibitemStop [0]{}%
\providecommand \bibitemNoStop [0]{.\EOS\space}%
\providecommand \EOS [0]{\spacefactor3000\relax}%
\providecommand \BibitemShut  [1]{\csname bibitem#1\endcsname}%
\let\auto@bib@innerbib\@empty
\bibitem [{\citenamefont {Snyder}\ and\ \citenamefont
  {Love}(2012)}]{snyder2012optical}%
  \BibitemOpen
  \bibfield  {author} {\bibinfo {author} {\bibfnamefont {Allan~W}\ \bibnamefont
  {Snyder}}\ and\ \bibinfo {author} {\bibfnamefont {John}\ \bibnamefont
  {Love}},\ }\href@noop {} {\emph {\bibinfo {title} {Optical waveguide
  theory}}}\ (\bibinfo  {publisher} {Springer Science \& Business Media},\
  \bibinfo {year} {2012})\BibitemShut {NoStop}%
\bibitem [{\citenamefont {Rademacher}\ \emph {et~al.}(2021)\citenamefont
  {Rademacher}, \citenamefont {Puttnam}, \citenamefont {Lu{\'\i}s},
  \citenamefont {Eriksson}, \citenamefont {Fontaine}, \citenamefont {Mazur},
  \citenamefont {Chen}, \citenamefont {Ryf}, \citenamefont {Neilson},
  \citenamefont {Sillard} \emph {et~al.}}]{rademacher2021peta}%
  \BibitemOpen
  \bibfield  {author} {\bibinfo {author} {\bibfnamefont {Georg}\ \bibnamefont
  {Rademacher}}, \bibinfo {author} {\bibfnamefont {Benjamin~J}\ \bibnamefont
  {Puttnam}}, \bibinfo {author} {\bibfnamefont {Ruben~S}\ \bibnamefont
  {Lu{\'\i}s}}, \bibinfo {author} {\bibfnamefont {Tobias~A}\ \bibnamefont
  {Eriksson}}, \bibinfo {author} {\bibfnamefont {Nicolas~K}\ \bibnamefont
  {Fontaine}}, \bibinfo {author} {\bibfnamefont {Mikael}\ \bibnamefont
  {Mazur}}, \bibinfo {author} {\bibfnamefont {Haoshuo}\ \bibnamefont {Chen}},
  \bibinfo {author} {\bibfnamefont {Roland}\ \bibnamefont {Ryf}}, \bibinfo
  {author} {\bibfnamefont {David~T}\ \bibnamefont {Neilson}}, \bibinfo {author}
  {\bibfnamefont {Pierre}\ \bibnamefont {Sillard}},  \emph {et~al.},\
  }\bibfield  {title} {\enquote {\bibinfo {title} {Peta-bit-per-second optical
  communications system using a standard cladding diameter 15-mode fiber},}\
  }\href@noop {} {\bibfield  {journal} {\bibinfo  {journal} {Nature
  Communications}\ }\textbf {\bibinfo {volume} {12}},\ \bibinfo {pages} {4238}
  (\bibinfo {year} {2021})}\BibitemShut {NoStop}%
\bibitem [{\citenamefont {Cao}\ \emph {et~al.}(2023)\citenamefont {Cao},
  \citenamefont {{\v{C}}i{\v{z}}m{\'a}r}, \citenamefont {Turtaev},
  \citenamefont {Tyc},\ and\ \citenamefont {Rotter}}]{cao2023controlling}%
  \BibitemOpen
  \bibfield  {author} {\bibinfo {author} {\bibfnamefont {Hui}\ \bibnamefont
  {Cao}}, \bibinfo {author} {\bibfnamefont {Tom{\'a}{\v{s}}}\ \bibnamefont
  {{\v{C}}i{\v{z}}m{\'a}r}}, \bibinfo {author} {\bibfnamefont {Sergey}\
  \bibnamefont {Turtaev}}, \bibinfo {author} {\bibfnamefont {Tom{\'a}{\v{s}}}\
  \bibnamefont {Tyc}}, \ and\ \bibinfo {author} {\bibfnamefont {Stefan}\
  \bibnamefont {Rotter}},\ }\bibfield  {title} {\enquote {\bibinfo {title}
  {Controlling light propagation in multimode fibers for imaging, spectroscopy
  and beyond},}\ }\href@noop {} {\bibfield  {journal} {\bibinfo  {journal}
  {arXiv preprint arXiv:2305.09623}\ } (\bibinfo {year} {2023})}\BibitemShut
  {NoStop}%
\bibitem [{\citenamefont {Gigan}\ \emph {et~al.}(2022)\citenamefont {Gigan},
  \citenamefont {Katz}, \citenamefont {De~Aguiar}, \citenamefont {Andresen},
  \citenamefont {Aubry}, \citenamefont {Bertolotti}, \citenamefont {Bossy},
  \citenamefont {Bouchet}, \citenamefont {Brake}, \citenamefont {Brasselet}
  \emph {et~al.}}]{gigan2022roadmap}%
  \BibitemOpen
  \bibfield  {author} {\bibinfo {author} {\bibfnamefont {Sylvain}\ \bibnamefont
  {Gigan}}, \bibinfo {author} {\bibfnamefont {Ori}\ \bibnamefont {Katz}},
  \bibinfo {author} {\bibfnamefont {Hilton~B}\ \bibnamefont {De~Aguiar}},
  \bibinfo {author} {\bibfnamefont {Esben~Ravn}\ \bibnamefont {Andresen}},
  \bibinfo {author} {\bibfnamefont {Alexandre}\ \bibnamefont {Aubry}}, \bibinfo
  {author} {\bibfnamefont {Jacopo}\ \bibnamefont {Bertolotti}}, \bibinfo
  {author} {\bibfnamefont {Emmanuel}\ \bibnamefont {Bossy}}, \bibinfo {author}
  {\bibfnamefont {Dorian}\ \bibnamefont {Bouchet}}, \bibinfo {author}
  {\bibfnamefont {Joshua}\ \bibnamefont {Brake}}, \bibinfo {author}
  {\bibfnamefont {Sophie}\ \bibnamefont {Brasselet}},  \emph {et~al.},\
  }\bibfield  {title} {\enquote {\bibinfo {title} {Roadmap on wavefront shaping
  and deep imaging in complex media},}\ }\href@noop {} {\bibfield  {journal}
  {\bibinfo  {journal} {Journal of Physics: Photonics}\ }\textbf {\bibinfo
  {volume} {4}},\ \bibinfo {pages} {042501} (\bibinfo {year}
  {2022})}\BibitemShut {NoStop}%
\bibitem [{\citenamefont {Stiburek}\ \emph {et~al.}(2023)\citenamefont
  {Stiburek}, \citenamefont {Ondr{\'a}{\v{c}}kov{\'a}}, \citenamefont
  {Tu{\v{c}}kov{\'a}}, \citenamefont {Turtaev}, \citenamefont {{\v{S}}iler},
  \citenamefont {Pik{\'a}lek}, \citenamefont {J{\'a}kl}, \citenamefont {Gomes},
  \citenamefont {Krej{\v{c}}{\'\i}}, \citenamefont {Kolb{\'a}bkov{\'a}} \emph
  {et~al.}}]{stibuurek2023110}%
  \BibitemOpen
  \bibfield  {author} {\bibinfo {author} {\bibfnamefont {Miroslav}\
  \bibnamefont {Stiburek}}, \bibinfo {author} {\bibfnamefont {Petra}\
  \bibnamefont {Ondr{\'a}{\v{c}}kov{\'a}}}, \bibinfo {author} {\bibfnamefont
  {Tereza}\ \bibnamefont {Tu{\v{c}}kov{\'a}}}, \bibinfo {author} {\bibfnamefont
  {Sergey}\ \bibnamefont {Turtaev}}, \bibinfo {author} {\bibfnamefont {Martin}\
  \bibnamefont {{\v{S}}iler}}, \bibinfo {author} {\bibfnamefont
  {Tom{\'a}{\v{s}}}\ \bibnamefont {Pik{\'a}lek}}, \bibinfo {author}
  {\bibfnamefont {Petr}\ \bibnamefont {J{\'a}kl}}, \bibinfo {author}
  {\bibfnamefont {Andr{\'e}}\ \bibnamefont {Gomes}}, \bibinfo {author}
  {\bibfnamefont {Jana}\ \bibnamefont {Krej{\v{c}}{\'\i}}}, \bibinfo {author}
  {\bibfnamefont {Petra}\ \bibnamefont {Kolb{\'a}bkov{\'a}}},  \emph {et~al.},\
  }\bibfield  {title} {\enquote {\bibinfo {title} {110 um thin endo-microscope
  for deep-brain in vivo observations of neuronal connectivity, activity and
  blood flow dynamics},}\ }\href@noop {} {\bibfield  {journal} {\bibinfo
  {journal} {Nature Communications}\ }\textbf {\bibinfo {volume} {14}},\
  \bibinfo {pages} {1897} (\bibinfo {year} {2023})}\BibitemShut {NoStop}%
\bibitem [{\citenamefont {Wen}\ \emph {et~al.}(2023)\citenamefont {Wen},
  \citenamefont {Dong}, \citenamefont {Deng}, \citenamefont {Pang},
  \citenamefont {Kaminski}, \citenamefont {Xu}, \citenamefont {Yan},
  \citenamefont {Wang}, \citenamefont {Liu}, \citenamefont {Tang} \emph
  {et~al.}}]{wen2023single}%
  \BibitemOpen
  \bibfield  {author} {\bibinfo {author} {\bibfnamefont {Zhong}\ \bibnamefont
  {Wen}}, \bibinfo {author} {\bibfnamefont {Zhenyu}\ \bibnamefont {Dong}},
  \bibinfo {author} {\bibfnamefont {Qilin}\ \bibnamefont {Deng}}, \bibinfo
  {author} {\bibfnamefont {Chenlei}\ \bibnamefont {Pang}}, \bibinfo {author}
  {\bibfnamefont {Clemens~F}\ \bibnamefont {Kaminski}}, \bibinfo {author}
  {\bibfnamefont {Xiaorong}\ \bibnamefont {Xu}}, \bibinfo {author}
  {\bibfnamefont {Huihui}\ \bibnamefont {Yan}}, \bibinfo {author}
  {\bibfnamefont {Liqiang}\ \bibnamefont {Wang}}, \bibinfo {author}
  {\bibfnamefont {Songguo}\ \bibnamefont {Liu}}, \bibinfo {author}
  {\bibfnamefont {Jianbin}\ \bibnamefont {Tang}},  \emph {et~al.},\ }\bibfield
  {title} {\enquote {\bibinfo {title} {Single multimode fibre for in vivo
  light-field-encoded endoscopic imaging},}\ }\href@noop {} {\bibfield
  {journal} {\bibinfo  {journal} {Nature Photonics}\ ,\ \bibinfo {pages}
  {1--9}} (\bibinfo {year} {2023})}\BibitemShut {NoStop}%
\bibitem [{\citenamefont {Richardson}\ \emph {et~al.}(2013)\citenamefont
  {Richardson}, \citenamefont {Fini},\ and\ \citenamefont
  {Nelson}}]{richardson2013space}%
  \BibitemOpen
  \bibfield  {author} {\bibinfo {author} {\bibfnamefont {David~J}\ \bibnamefont
  {Richardson}}, \bibinfo {author} {\bibfnamefont {John~M}\ \bibnamefont
  {Fini}}, \ and\ \bibinfo {author} {\bibfnamefont {Lynn~E}\ \bibnamefont
  {Nelson}},\ }\bibfield  {title} {\enquote {\bibinfo {title} {Space-division
  multiplexing in optical fibres},}\ }\href@noop {} {\bibfield  {journal}
  {\bibinfo  {journal} {Nature photonics}\ }\textbf {\bibinfo {volume} {7}},\
  \bibinfo {pages} {354--362} (\bibinfo {year} {2013})}\BibitemShut {NoStop}%
\bibitem [{\citenamefont {Cristiani}\ \emph {et~al.}(2022)\citenamefont
  {Cristiani}, \citenamefont {Lacava}, \citenamefont {Rademacher},
  \citenamefont {Puttnam}, \citenamefont {Lu{\`\i}s}, \citenamefont
  {Antonelli}, \citenamefont {Mecozzi}, \citenamefont {Shtaif}, \citenamefont
  {Cozzolino}, \citenamefont {Bacco} \emph {et~al.}}]{cristiani2022roadmap}%
  \BibitemOpen
  \bibfield  {author} {\bibinfo {author} {\bibfnamefont {Ilaria}\ \bibnamefont
  {Cristiani}}, \bibinfo {author} {\bibfnamefont {Cosimo}\ \bibnamefont
  {Lacava}}, \bibinfo {author} {\bibfnamefont {Georg}\ \bibnamefont
  {Rademacher}}, \bibinfo {author} {\bibfnamefont {Benjamin~J}\ \bibnamefont
  {Puttnam}}, \bibinfo {author} {\bibfnamefont {Ruben~S}\ \bibnamefont
  {Lu{\`\i}s}}, \bibinfo {author} {\bibfnamefont {Cristian}\ \bibnamefont
  {Antonelli}}, \bibinfo {author} {\bibfnamefont {Antonio}\ \bibnamefont
  {Mecozzi}}, \bibinfo {author} {\bibfnamefont {Mark}\ \bibnamefont {Shtaif}},
  \bibinfo {author} {\bibfnamefont {Daniele}\ \bibnamefont {Cozzolino}},
  \bibinfo {author} {\bibfnamefont {Davide}\ \bibnamefont {Bacco}},  \emph
  {et~al.},\ }\bibfield  {title} {\enquote {\bibinfo {title} {Roadmap on
  multimode photonics},}\ }\href@noop {} {\bibfield  {journal} {\bibinfo
  {journal} {Journal of Optics}\ }\textbf {\bibinfo {volume} {24}},\ \bibinfo
  {pages} {083001} (\bibinfo {year} {2022})}\BibitemShut {NoStop}%
\bibitem [{\citenamefont {Rahmani}\ \emph {et~al.}(2022)\citenamefont
  {Rahmani}, \citenamefont {Oguz}, \citenamefont {Tegin}, \citenamefont
  {Hsieh}, \citenamefont {Psaltis},\ and\ \citenamefont
  {Moser}}]{rahmani2022learning}%
  \BibitemOpen
  \bibfield  {author} {\bibinfo {author} {\bibfnamefont {Babak}\ \bibnamefont
  {Rahmani}}, \bibinfo {author} {\bibfnamefont {Ilker}\ \bibnamefont {Oguz}},
  \bibinfo {author} {\bibfnamefont {Ugur}\ \bibnamefont {Tegin}}, \bibinfo
  {author} {\bibfnamefont {Jih-liang}\ \bibnamefont {Hsieh}}, \bibinfo {author}
  {\bibfnamefont {Demetri}\ \bibnamefont {Psaltis}}, \ and\ \bibinfo {author}
  {\bibfnamefont {Christophe}\ \bibnamefont {Moser}},\ }\bibfield  {title}
  {\enquote {\bibinfo {title} {Learning to image and compute with multimode
  optical fibers},}\ }\href@noop {} {\bibfield  {journal} {\bibinfo  {journal}
  {Nanophotonics}\ }\textbf {\bibinfo {volume} {11}},\ \bibinfo {pages}
  {1071--1082} (\bibinfo {year} {2022})}\BibitemShut {NoStop}%
\bibitem [{\citenamefont {Te{\u{g}}in}\ \emph {et~al.}(2021)\citenamefont
  {Te{\u{g}}in}, \citenamefont {Y{\i}ld{\i}r{\i}m}, \citenamefont {O{\u{g}}uz},
  \citenamefont {Moser},\ and\ \citenamefont {Psaltis}}]{teugin2021scalable}%
  \BibitemOpen
  \bibfield  {author} {\bibinfo {author} {\bibfnamefont {U{\u{g}}ur}\
  \bibnamefont {Te{\u{g}}in}}, \bibinfo {author} {\bibfnamefont {Mustafa}\
  \bibnamefont {Y{\i}ld{\i}r{\i}m}}, \bibinfo {author} {\bibfnamefont
  {{\.I}lker}\ \bibnamefont {O{\u{g}}uz}}, \bibinfo {author} {\bibfnamefont
  {Christophe}\ \bibnamefont {Moser}}, \ and\ \bibinfo {author} {\bibfnamefont
  {Demetri}\ \bibnamefont {Psaltis}},\ }\bibfield  {title} {\enquote {\bibinfo
  {title} {Scalable optical learning operator},}\ }\href@noop {} {\bibfield
  {journal} {\bibinfo  {journal} {Nature Computational Science}\ }\textbf
  {\bibinfo {volume} {1}},\ \bibinfo {pages} {542--549} (\bibinfo {year}
  {2021})}\BibitemShut {NoStop}%
\bibitem [{\citenamefont {Leedumrongwatthanakun}\ \emph
  {et~al.}(2020)\citenamefont {Leedumrongwatthanakun}, \citenamefont
  {Innocenti}, \citenamefont {Defienne}, \citenamefont {Juffmann},
  \citenamefont {Ferraro}, \citenamefont {Paternostro},\ and\ \citenamefont
  {Gigan}}]{leedumrongwatthanakun2020programmable}%
  \BibitemOpen
  \bibfield  {author} {\bibinfo {author} {\bibfnamefont {Saroch}\ \bibnamefont
  {Leedumrongwatthanakun}}, \bibinfo {author} {\bibfnamefont {Luca}\
  \bibnamefont {Innocenti}}, \bibinfo {author} {\bibfnamefont {Hugo}\
  \bibnamefont {Defienne}}, \bibinfo {author} {\bibfnamefont {Thomas}\
  \bibnamefont {Juffmann}}, \bibinfo {author} {\bibfnamefont {Alessandro}\
  \bibnamefont {Ferraro}}, \bibinfo {author} {\bibfnamefont {Mauro}\
  \bibnamefont {Paternostro}}, \ and\ \bibinfo {author} {\bibfnamefont
  {Sylvain}\ \bibnamefont {Gigan}},\ }\bibfield  {title} {\enquote {\bibinfo
  {title} {Programmable linear quantum networks with a multimode fibre},}\
  }\href@noop {} {\bibfield  {journal} {\bibinfo  {journal} {Nature Photonics}\
  }\textbf {\bibinfo {volume} {14}},\ \bibinfo {pages} {139--142} (\bibinfo
  {year} {2020})}\BibitemShut {NoStop}%
\bibitem [{\citenamefont {Popoff}\ \emph {et~al.}(2010)\citenamefont {Popoff},
  \citenamefont {Lerosey}, \citenamefont {Carminati}, \citenamefont {Fink},
  \citenamefont {Boccara},\ and\ \citenamefont {Gigan}}]{popoff2010measuring}%
  \BibitemOpen
  \bibfield  {author} {\bibinfo {author} {\bibfnamefont {S{\'e}bastien~M}\
  \bibnamefont {Popoff}}, \bibinfo {author} {\bibfnamefont {Geoffroy}\
  \bibnamefont {Lerosey}}, \bibinfo {author} {\bibfnamefont {R{\'e}mi}\
  \bibnamefont {Carminati}}, \bibinfo {author} {\bibfnamefont {Mathias}\
  \bibnamefont {Fink}}, \bibinfo {author} {\bibfnamefont {Albert~Claude}\
  \bibnamefont {Boccara}}, \ and\ \bibinfo {author} {\bibfnamefont {Sylvain}\
  \bibnamefont {Gigan}},\ }\bibfield  {title} {\enquote {\bibinfo {title}
  {Measuring the transmission matrix in optics: an approach to the study and
  control of light propagation in disordered media},}\ }\href@noop {}
  {\bibfield  {journal} {\bibinfo  {journal} {Physical review letters}\
  }\textbf {\bibinfo {volume} {104}},\ \bibinfo {pages} {100601} (\bibinfo
  {year} {2010})}\BibitemShut {NoStop}%
\bibitem [{\citenamefont {Choi}\ \emph {et~al.}(2012)\citenamefont {Choi},
  \citenamefont {Yoon}, \citenamefont {Kim}, \citenamefont {Yang},
  \citenamefont {Fang-Yen}, \citenamefont {Dasari}, \citenamefont {Lee},\ and\
  \citenamefont {Choi}}]{choi2012scanner}%
  \BibitemOpen
  \bibfield  {author} {\bibinfo {author} {\bibfnamefont {Youngwoon}\
  \bibnamefont {Choi}}, \bibinfo {author} {\bibfnamefont {Changhyeong}\
  \bibnamefont {Yoon}}, \bibinfo {author} {\bibfnamefont {Moonseok}\
  \bibnamefont {Kim}}, \bibinfo {author} {\bibfnamefont {Taeseok~Daniel}\
  \bibnamefont {Yang}}, \bibinfo {author} {\bibfnamefont {Christopher}\
  \bibnamefont {Fang-Yen}}, \bibinfo {author} {\bibfnamefont {Ramachandra~R}\
  \bibnamefont {Dasari}}, \bibinfo {author} {\bibfnamefont {Kyoung~Jin}\
  \bibnamefont {Lee}}, \ and\ \bibinfo {author} {\bibfnamefont {Wonshik}\
  \bibnamefont {Choi}},\ }\bibfield  {title} {\enquote {\bibinfo {title}
  {Scanner-free and wide-field endoscopic imaging by using a single multimode
  optical fiber},}\ }\href@noop {} {\bibfield  {journal} {\bibinfo  {journal}
  {Physical review letters}\ }\textbf {\bibinfo {volume} {109}},\ \bibinfo
  {pages} {203901} (\bibinfo {year} {2012})}\BibitemShut {NoStop}%
\bibitem [{\citenamefont {Lee}\ \emph {et~al.}(2022)\citenamefont {Lee},
  \citenamefont {Parot}, \citenamefont {Bouma},\ and\ \citenamefont
  {Villiger}}]{lee2022confocal}%
  \BibitemOpen
  \bibfield  {author} {\bibinfo {author} {\bibfnamefont {Szu-Yu}\ \bibnamefont
  {Lee}}, \bibinfo {author} {\bibfnamefont {Vicente~J}\ \bibnamefont {Parot}},
  \bibinfo {author} {\bibfnamefont {Brett~E}\ \bibnamefont {Bouma}}, \ and\
  \bibinfo {author} {\bibfnamefont {Martin}\ \bibnamefont {Villiger}},\
  }\bibfield  {title} {\enquote {\bibinfo {title} {Confocal 3d reflectance
  imaging through multimode fiber without wavefront shaping},}\ }\href@noop {}
  {\bibfield  {journal} {\bibinfo  {journal} {Optica}\ }\textbf {\bibinfo
  {volume} {9}},\ \bibinfo {pages} {112--120} (\bibinfo {year}
  {2022})}\BibitemShut {NoStop}%
\bibitem [{\citenamefont {Shah}\ \emph {et~al.}(2005)\citenamefont {Shah},
  \citenamefont {Hsu}, \citenamefont {Tarighat}, \citenamefont {Sayed},\ and\
  \citenamefont {Jalali}}]{shah2005coherent}%
  \BibitemOpen
  \bibfield  {author} {\bibinfo {author} {\bibfnamefont {Akhil~R}\ \bibnamefont
  {Shah}}, \bibinfo {author} {\bibfnamefont {Rick~CJ}\ \bibnamefont {Hsu}},
  \bibinfo {author} {\bibfnamefont {Alireza}\ \bibnamefont {Tarighat}},
  \bibinfo {author} {\bibfnamefont {Ali~H}\ \bibnamefont {Sayed}}, \ and\
  \bibinfo {author} {\bibfnamefont {Bahram}\ \bibnamefont {Jalali}},\
  }\bibfield  {title} {\enquote {\bibinfo {title} {Coherent optical mimo
  (comimo)},}\ }\href@noop {} {\bibfield  {journal} {\bibinfo  {journal}
  {Journal of lightwave technology}\ }\textbf {\bibinfo {volume} {23}},\
  \bibinfo {pages} {2410--2419} (\bibinfo {year} {2005})}\BibitemShut {NoStop}%
\bibitem [{\citenamefont {Rademacher}\ \emph {et~al.}(2022)\citenamefont
  {Rademacher}, \citenamefont {Lu{\'\i}s}, \citenamefont {Puttnam},
  \citenamefont {Fontaine}, \citenamefont {Mazur}, \citenamefont {Chen},
  \citenamefont {Ryf}, \citenamefont {Neilson}, \citenamefont {Dahl},
  \citenamefont {Carpenter} \emph {et~al.}}]{rademacher20221}%
  \BibitemOpen
  \bibfield  {author} {\bibinfo {author} {\bibfnamefont {Georg}\ \bibnamefont
  {Rademacher}}, \bibinfo {author} {\bibfnamefont {Ruben~S}\ \bibnamefont
  {Lu{\'\i}s}}, \bibinfo {author} {\bibfnamefont {Benjamin~J}\ \bibnamefont
  {Puttnam}}, \bibinfo {author} {\bibfnamefont {Nicolas~K}\ \bibnamefont
  {Fontaine}}, \bibinfo {author} {\bibfnamefont {Mikael}\ \bibnamefont
  {Mazur}}, \bibinfo {author} {\bibfnamefont {Haoshuo}\ \bibnamefont {Chen}},
  \bibinfo {author} {\bibfnamefont {Roland}\ \bibnamefont {Ryf}}, \bibinfo
  {author} {\bibfnamefont {David~T}\ \bibnamefont {Neilson}}, \bibinfo {author}
  {\bibfnamefont {Daniel}\ \bibnamefont {Dahl}}, \bibinfo {author}
  {\bibfnamefont {Joel}\ \bibnamefont {Carpenter}},  \emph {et~al.},\
  }\bibfield  {title} {\enquote {\bibinfo {title} {1.53 peta-bit/s c-band
  transmission in a 55-mode fiber},}\ }in\ \href@noop {} {\emph {\bibinfo
  {booktitle} {2022 European Conference on Optical Communication (ECOC)}}}\
  (\bibinfo {organization} {IEEE},\ \bibinfo {year} {2022})\ pp.\ \bibinfo
  {pages} {1--4}\BibitemShut {NoStop}%
\bibitem [{\citenamefont {{\v{C}}i{\v{z}}m{\'a}r}\ and\ \citenamefont
  {Dholakia}(2012)}]{vcivzmar2012exploiting}%
  \BibitemOpen
  \bibfield  {author} {\bibinfo {author} {\bibfnamefont {Tom{\'a}{\v{s}}}\
  \bibnamefont {{\v{C}}i{\v{z}}m{\'a}r}}\ and\ \bibinfo {author} {\bibfnamefont
  {Kishan}\ \bibnamefont {Dholakia}},\ }\bibfield  {title} {\enquote {\bibinfo
  {title} {Exploiting multimode waveguides for pure fibre-based imaging},}\
  }\href@noop {} {\bibfield  {journal} {\bibinfo  {journal} {Nature
  communications}\ }\textbf {\bibinfo {volume} {3}},\ \bibinfo {pages} {1--9}
  (\bibinfo {year} {2012})}\BibitemShut {NoStop}%
\bibitem [{\citenamefont {Vellekoop}\ and\ \citenamefont
  {Mosk}(2007)}]{vellekoop2007focusing}%
  \BibitemOpen
  \bibfield  {author} {\bibinfo {author} {\bibfnamefont {Ivo~M}\ \bibnamefont
  {Vellekoop}}\ and\ \bibinfo {author} {\bibfnamefont {AP}~\bibnamefont
  {Mosk}},\ }\bibfield  {title} {\enquote {\bibinfo {title} {Focusing coherent
  light through opaque strongly scattering media},}\ }\href@noop {} {\bibfield
  {journal} {\bibinfo  {journal} {Optics letters}\ }\textbf {\bibinfo {volume}
  {32}},\ \bibinfo {pages} {2309--2311} (\bibinfo {year} {2007})}\BibitemShut
  {NoStop}%
\bibitem [{\citenamefont {Papadopoulos}\ \emph {et~al.}(2012)\citenamefont
  {Papadopoulos}, \citenamefont {Farahi}, \citenamefont {Moser},\ and\
  \citenamefont {Psaltis}}]{papadopoulos2012focusing}%
  \BibitemOpen
  \bibfield  {author} {\bibinfo {author} {\bibfnamefont {Ioannis~N}\
  \bibnamefont {Papadopoulos}}, \bibinfo {author} {\bibfnamefont {Salma}\
  \bibnamefont {Farahi}}, \bibinfo {author} {\bibfnamefont {Christophe}\
  \bibnamefont {Moser}}, \ and\ \bibinfo {author} {\bibfnamefont {Demetri}\
  \bibnamefont {Psaltis}},\ }\bibfield  {title} {\enquote {\bibinfo {title}
  {Focusing and scanning light through a multimode optical fiber using digital
  phase conjugation},}\ }\href@noop {} {\bibfield  {journal} {\bibinfo
  {journal} {Optics express}\ }\textbf {\bibinfo {volume} {20}},\ \bibinfo
  {pages} {10583--10590} (\bibinfo {year} {2012})}\BibitemShut {NoStop}%
\bibitem [{\citenamefont {Loterie}\ \emph {et~al.}(2015)\citenamefont
  {Loterie}, \citenamefont {Farahi}, \citenamefont {Papadopoulos},
  \citenamefont {Goy}, \citenamefont {Psaltis},\ and\ \citenamefont
  {Moser}}]{loterie2015digital}%
  \BibitemOpen
  \bibfield  {author} {\bibinfo {author} {\bibfnamefont {Damien}\ \bibnamefont
  {Loterie}}, \bibinfo {author} {\bibfnamefont {Salma}\ \bibnamefont {Farahi}},
  \bibinfo {author} {\bibfnamefont {Ioannis}\ \bibnamefont {Papadopoulos}},
  \bibinfo {author} {\bibfnamefont {Alexandre}\ \bibnamefont {Goy}}, \bibinfo
  {author} {\bibfnamefont {Demetri}\ \bibnamefont {Psaltis}}, \ and\ \bibinfo
  {author} {\bibfnamefont {Christophe}\ \bibnamefont {Moser}},\ }\bibfield
  {title} {\enquote {\bibinfo {title} {Digital confocal microscopy through a
  multimode fiber},}\ }\href@noop {} {\bibfield  {journal} {\bibinfo  {journal}
  {Optics express}\ }\textbf {\bibinfo {volume} {23}},\ \bibinfo {pages}
  {23845--23858} (\bibinfo {year} {2015})}\BibitemShut {NoStop}%
\bibitem [{\citenamefont {Gusachenko}\ \emph {et~al.}(2017)\citenamefont
  {Gusachenko}, \citenamefont {Chen},\ and\ \citenamefont
  {Dholakia}}]{gusachenko2017raman}%
  \BibitemOpen
  \bibfield  {author} {\bibinfo {author} {\bibfnamefont {Ivan}\ \bibnamefont
  {Gusachenko}}, \bibinfo {author} {\bibfnamefont {Mingzhou}\ \bibnamefont
  {Chen}}, \ and\ \bibinfo {author} {\bibfnamefont {Kishan}\ \bibnamefont
  {Dholakia}},\ }\bibfield  {title} {\enquote {\bibinfo {title} {Raman imaging
  through a single multimode fibre},}\ }\href@noop {} {\bibfield  {journal}
  {\bibinfo  {journal} {Optics express}\ }\textbf {\bibinfo {volume} {25}},\
  \bibinfo {pages} {13782--13798} (\bibinfo {year} {2017})}\BibitemShut
  {NoStop}%
\bibitem [{\citenamefont {Turtaev}\ \emph {et~al.}(2018)\citenamefont
  {Turtaev}, \citenamefont {Leite}, \citenamefont {Altwegg-Boussac},
  \citenamefont {Pakan}, \citenamefont {Rochefort},\ and\ \citenamefont
  {{\v{C}}i{\v{z}}m{\'a}r}}]{turtaev2018high}%
  \BibitemOpen
  \bibfield  {author} {\bibinfo {author} {\bibfnamefont {Sergey}\ \bibnamefont
  {Turtaev}}, \bibinfo {author} {\bibfnamefont {Ivo~T}\ \bibnamefont {Leite}},
  \bibinfo {author} {\bibfnamefont {Tristan}\ \bibnamefont {Altwegg-Boussac}},
  \bibinfo {author} {\bibfnamefont {Janelle~MP}\ \bibnamefont {Pakan}},
  \bibinfo {author} {\bibfnamefont {Nathalie~L}\ \bibnamefont {Rochefort}}, \
  and\ \bibinfo {author} {\bibfnamefont {Tom{\'a}{\v{s}}}\ \bibnamefont
  {{\v{C}}i{\v{z}}m{\'a}r}},\ }\bibfield  {title} {\enquote {\bibinfo {title}
  {High-fidelity multimode fibre-based endoscopy for deep brain in vivo
  imaging},}\ }\href@noop {} {\bibfield  {journal} {\bibinfo  {journal} {Light:
  Science \& Applications}\ }\textbf {\bibinfo {volume} {7}},\ \bibinfo {pages}
  {1--8} (\bibinfo {year} {2018})}\BibitemShut {NoStop}%
\bibitem [{\citenamefont {Ohayon}\ \emph {et~al.}(2018)\citenamefont {Ohayon},
  \citenamefont {Caravaca-Aguirre}, \citenamefont {Piestun},\ and\
  \citenamefont {DiCarlo}}]{ohayon2018minimally}%
  \BibitemOpen
  \bibfield  {author} {\bibinfo {author} {\bibfnamefont {Shay}\ \bibnamefont
  {Ohayon}}, \bibinfo {author} {\bibfnamefont {Antonio}\ \bibnamefont
  {Caravaca-Aguirre}}, \bibinfo {author} {\bibfnamefont {Rafael}\ \bibnamefont
  {Piestun}}, \ and\ \bibinfo {author} {\bibfnamefont {James~J}\ \bibnamefont
  {DiCarlo}},\ }\bibfield  {title} {\enquote {\bibinfo {title} {Minimally
  invasive multimode optical fiber microendoscope for deep brain fluorescence
  imaging},}\ }\href@noop {} {\bibfield  {journal} {\bibinfo  {journal}
  {Biomedical optics express}\ }\textbf {\bibinfo {volume} {9}},\ \bibinfo
  {pages} {1492--1509} (\bibinfo {year} {2018})}\BibitemShut {NoStop}%
\bibitem [{\citenamefont {Tr{\"a}g{\aa}rdh}\ \emph {et~al.}(2019)\citenamefont
  {Tr{\"a}g{\aa}rdh}, \citenamefont {Pik{\'a}lek}, \citenamefont
  {{\v{S}}er{\'y}}, \citenamefont {Meyer}, \citenamefont {Popp},\ and\
  \citenamefont {{\v{C}}i{\v{z}}m{\'a}r}}]{tragaardh2019label}%
  \BibitemOpen
  \bibfield  {author} {\bibinfo {author} {\bibfnamefont {Johanna}\ \bibnamefont
  {Tr{\"a}g{\aa}rdh}}, \bibinfo {author} {\bibfnamefont {Tom{\'a}{\v{s}}}\
  \bibnamefont {Pik{\'a}lek}}, \bibinfo {author} {\bibfnamefont {Mojm{\'\i}r}\
  \bibnamefont {{\v{S}}er{\'y}}}, \bibinfo {author} {\bibfnamefont {Tobias}\
  \bibnamefont {Meyer}}, \bibinfo {author} {\bibfnamefont {J{\"u}rgen}\
  \bibnamefont {Popp}}, \ and\ \bibinfo {author} {\bibfnamefont
  {Tom{\'a}{\v{s}}}\ \bibnamefont {{\v{C}}i{\v{z}}m{\'a}r}},\ }\bibfield
  {title} {\enquote {\bibinfo {title} {Label-free cars microscopy through a
  multimode fiber endoscope},}\ }\href@noop {} {\bibfield  {journal} {\bibinfo
  {journal} {Optics express}\ }\textbf {\bibinfo {volume} {27}},\ \bibinfo
  {pages} {30055--30066} (\bibinfo {year} {2019})}\BibitemShut {NoStop}%
\bibitem [{\citenamefont {Cifuentes}\ \emph {et~al.}(2021)\citenamefont
  {Cifuentes}, \citenamefont {Pik{\'a}lek}, \citenamefont
  {Ondr{\'a}{\v{c}}kov{\'a}}, \citenamefont {Amezcua-Correa}, \citenamefont
  {Antonio-Lopez}, \citenamefont {{\v{C}}i{\v{z}}m{\'a}r},\ and\ \citenamefont
  {Tr{\"a}g{\aa}rdh}}]{cifuentes2021polarization}%
  \BibitemOpen
  \bibfield  {author} {\bibinfo {author} {\bibfnamefont {Angel}\ \bibnamefont
  {Cifuentes}}, \bibinfo {author} {\bibfnamefont {Tom{\'a}{\v{s}}}\
  \bibnamefont {Pik{\'a}lek}}, \bibinfo {author} {\bibfnamefont {Petra}\
  \bibnamefont {Ondr{\'a}{\v{c}}kov{\'a}}}, \bibinfo {author} {\bibfnamefont
  {Rodrigo}\ \bibnamefont {Amezcua-Correa}}, \bibinfo {author} {\bibfnamefont
  {Jos{\'e}~Enrique}\ \bibnamefont {Antonio-Lopez}}, \bibinfo {author}
  {\bibfnamefont {Tom{\'a}{\v{s}}}\ \bibnamefont {{\v{C}}i{\v{z}}m{\'a}r}}, \
  and\ \bibinfo {author} {\bibfnamefont {Johanna}\ \bibnamefont
  {Tr{\"a}g{\aa}rdh}},\ }\bibfield  {title} {\enquote {\bibinfo {title}
  {Polarization-resolved second-harmonic generation imaging through a multimode
  fiber},}\ }\href@noop {} {\bibfield  {journal} {\bibinfo  {journal} {Optica}\
  }\textbf {\bibinfo {volume} {8}},\ \bibinfo {pages} {1065--1074} (\bibinfo
  {year} {2021})}\BibitemShut {NoStop}%
\bibitem [{\citenamefont {Leite}\ \emph {et~al.}(2021)\citenamefont {Leite},
  \citenamefont {Turtaev}, \citenamefont {Boonzajer~Flaes},\ and\ \citenamefont
  {{\v{C}}i{\v{z}}m{\'a}r}}]{leite2021observing}%
  \BibitemOpen
  \bibfield  {author} {\bibinfo {author} {\bibfnamefont {Ivo~T}\ \bibnamefont
  {Leite}}, \bibinfo {author} {\bibfnamefont {Sergey}\ \bibnamefont {Turtaev}},
  \bibinfo {author} {\bibfnamefont {Dirk~E}\ \bibnamefont {Boonzajer~Flaes}}, \
  and\ \bibinfo {author} {\bibfnamefont {Tom{\'a}{\v{s}}}\ \bibnamefont
  {{\v{C}}i{\v{z}}m{\'a}r}},\ }\bibfield  {title} {\enquote {\bibinfo {title}
  {Observing distant objects with a multimode fiber-based holographic
  endoscope},}\ }\href@noop {} {\bibfield  {journal} {\bibinfo  {journal} {APL
  Photonics}\ }\textbf {\bibinfo {volume} {6}},\ \bibinfo {pages} {036112}
  (\bibinfo {year} {2021})}\BibitemShut {NoStop}%
\bibitem [{\citenamefont {Stellinga}\ \emph {et~al.}(2021)\citenamefont
  {Stellinga}, \citenamefont {Phillips}, \citenamefont {Mekhail}, \citenamefont
  {Selyem}, \citenamefont {Turtaev}, \citenamefont {{\v{C}}i{\v{z}}m{\'a}r},\
  and\ \citenamefont {Padgett}}]{stellinga2021time}%
  \BibitemOpen
  \bibfield  {author} {\bibinfo {author} {\bibfnamefont {Daan}\ \bibnamefont
  {Stellinga}}, \bibinfo {author} {\bibfnamefont {David~B}\ \bibnamefont
  {Phillips}}, \bibinfo {author} {\bibfnamefont {Simon~Peter}\ \bibnamefont
  {Mekhail}}, \bibinfo {author} {\bibfnamefont {Adam}\ \bibnamefont {Selyem}},
  \bibinfo {author} {\bibfnamefont {Sergey}\ \bibnamefont {Turtaev}}, \bibinfo
  {author} {\bibfnamefont {Tom{\'a}{\v{s}}}\ \bibnamefont
  {{\v{C}}i{\v{z}}m{\'a}r}}, \ and\ \bibinfo {author} {\bibfnamefont {Miles~J}\
  \bibnamefont {Padgett}},\ }\bibfield  {title} {\enquote {\bibinfo {title}
  {Time-of-flight 3d imaging through multimode optical fibers},}\ }\href@noop
  {} {\bibfield  {journal} {\bibinfo  {journal} {Science}\ }\textbf {\bibinfo
  {volume} {374}},\ \bibinfo {pages} {1395--1399} (\bibinfo {year}
  {2021})}\BibitemShut {NoStop}%
\bibitem [{\citenamefont {Betzig}\ \emph {et~al.}(2006)\citenamefont {Betzig},
  \citenamefont {Patterson}, \citenamefont {Sougrat}, \citenamefont
  {Lindwasser}, \citenamefont {Olenych}, \citenamefont {Bonifacino},
  \citenamefont {Davidson}, \citenamefont {Lippincott-Schwartz},\ and\
  \citenamefont {Hess}}]{betzig2006imaging}%
  \BibitemOpen
  \bibfield  {author} {\bibinfo {author} {\bibfnamefont {Eric}\ \bibnamefont
  {Betzig}}, \bibinfo {author} {\bibfnamefont {George~H}\ \bibnamefont
  {Patterson}}, \bibinfo {author} {\bibfnamefont {Rachid}\ \bibnamefont
  {Sougrat}}, \bibinfo {author} {\bibfnamefont {O~Wolf}\ \bibnamefont
  {Lindwasser}}, \bibinfo {author} {\bibfnamefont {Scott}\ \bibnamefont
  {Olenych}}, \bibinfo {author} {\bibfnamefont {Juan~S}\ \bibnamefont
  {Bonifacino}}, \bibinfo {author} {\bibfnamefont {Michael~W}\ \bibnamefont
  {Davidson}}, \bibinfo {author} {\bibfnamefont {Jennifer}\ \bibnamefont
  {Lippincott-Schwartz}}, \ and\ \bibinfo {author} {\bibfnamefont {Harald~F}\
  \bibnamefont {Hess}},\ }\bibfield  {title} {\enquote {\bibinfo {title}
  {Imaging intracellular fluorescent proteins at nanometer resolution},}\
  }\href@noop {} {\bibfield  {journal} {\bibinfo  {journal} {science}\ }\textbf
  {\bibinfo {volume} {313}},\ \bibinfo {pages} {1642--1645} (\bibinfo {year}
  {2006})}\BibitemShut {NoStop}%
\bibitem [{\citenamefont {Rust}\ \emph {et~al.}(2006)\citenamefont {Rust},
  \citenamefont {Bates},\ and\ \citenamefont {Zhuang}}]{rust2006sub}%
  \BibitemOpen
  \bibfield  {author} {\bibinfo {author} {\bibfnamefont {Michael~J}\
  \bibnamefont {Rust}}, \bibinfo {author} {\bibfnamefont {Mark}\ \bibnamefont
  {Bates}}, \ and\ \bibinfo {author} {\bibfnamefont {Xiaowei}\ \bibnamefont
  {Zhuang}},\ }\bibfield  {title} {\enquote {\bibinfo {title}
  {Sub-diffraction-limit imaging by stochastic optical reconstruction
  microscopy (storm)},}\ }\href@noop {} {\bibfield  {journal} {\bibinfo
  {journal} {Nature methods}\ }\textbf {\bibinfo {volume} {3}},\ \bibinfo
  {pages} {793--796} (\bibinfo {year} {2006})}\BibitemShut {NoStop}%
\bibitem [{\citenamefont {Liu}\ \emph {et~al.}(2022)\citenamefont {Liu},
  \citenamefont {Zhang}, \citenamefont {Liu}, \citenamefont {Lin},
  \citenamefont {Li}, \citenamefont {Lin}, \citenamefont {Zhang}, \citenamefont
  {Huang}, \citenamefont {Mo}, \citenamefont {Shen} \emph {et~al.}}]{liu20221}%
  \BibitemOpen
  \bibfield  {author} {\bibinfo {author} {\bibfnamefont {Junyi}\ \bibnamefont
  {Liu}}, \bibinfo {author} {\bibfnamefont {Jingxing}\ \bibnamefont {Zhang}},
  \bibinfo {author} {\bibfnamefont {Jie}\ \bibnamefont {Liu}}, \bibinfo
  {author} {\bibfnamefont {Zhenrui}\ \bibnamefont {Lin}}, \bibinfo {author}
  {\bibfnamefont {Zhenhua}\ \bibnamefont {Li}}, \bibinfo {author}
  {\bibfnamefont {Zhongzheng}\ \bibnamefont {Lin}}, \bibinfo {author}
  {\bibfnamefont {Junwei}\ \bibnamefont {Zhang}}, \bibinfo {author}
  {\bibfnamefont {Cong}\ \bibnamefont {Huang}}, \bibinfo {author}
  {\bibfnamefont {Shuqi}\ \bibnamefont {Mo}}, \bibinfo {author} {\bibfnamefont
  {Lei}\ \bibnamefont {Shen}},  \emph {et~al.},\ }\bibfield  {title} {\enquote
  {\bibinfo {title} {1-pbps orbital angular momentum fibre-optic
  transmission},}\ }\href@noop {} {\bibfield  {journal} {\bibinfo  {journal}
  {Light: Science \& Applications}\ }\textbf {\bibinfo {volume} {11}},\
  \bibinfo {pages} {202} (\bibinfo {year} {2022})}\BibitemShut {NoStop}%
\bibitem [{\citenamefont {Doerr}(2011)}]{doerr2011proposed}%
  \BibitemOpen
  \bibfield  {author} {\bibinfo {author} {\bibfnamefont {Christopher~R}\
  \bibnamefont {Doerr}},\ }\bibfield  {title} {\enquote {\bibinfo {title}
  {Proposed architecture for mimo optical demultiplexing using photonic
  integration},}\ }\href@noop {} {\bibfield  {journal} {\bibinfo  {journal}
  {IEEE Photonics Technology Letters}\ }\textbf {\bibinfo {volume} {23}},\
  \bibinfo {pages} {1573--1575} (\bibinfo {year} {2011})}\BibitemShut {NoStop}%
\bibitem [{\citenamefont {Tanomura}\ \emph {et~al.}(2020)\citenamefont
  {Tanomura}, \citenamefont {Tang}, \citenamefont {Suganuma}, \citenamefont
  {Okawa}, \citenamefont {Kato}, \citenamefont {Tanemura},\ and\ \citenamefont
  {Nakano}}]{tanomura2020monolithic}%
  \BibitemOpen
  \bibfield  {author} {\bibinfo {author} {\bibfnamefont {Ryota}\ \bibnamefont
  {Tanomura}}, \bibinfo {author} {\bibfnamefont {Rui}\ \bibnamefont {Tang}},
  \bibinfo {author} {\bibfnamefont {Takahiro}\ \bibnamefont {Suganuma}},
  \bibinfo {author} {\bibfnamefont {Kosuke}\ \bibnamefont {Okawa}}, \bibinfo
  {author} {\bibfnamefont {Eisaku}\ \bibnamefont {Kato}}, \bibinfo {author}
  {\bibfnamefont {Takuo}\ \bibnamefont {Tanemura}}, \ and\ \bibinfo {author}
  {\bibfnamefont {Yoshiaki}\ \bibnamefont {Nakano}},\ }\bibfield  {title}
  {\enquote {\bibinfo {title} {Monolithic inp optical unitary converter based
  on multi-plane light conversion},}\ }\href@noop {} {\bibfield  {journal}
  {\bibinfo  {journal} {Optics Express}\ }\textbf {\bibinfo {volume} {28}},\
  \bibinfo {pages} {25392--25399} (\bibinfo {year} {2020})}\BibitemShut
  {NoStop}%
\bibitem [{\citenamefont {B{\=u}tait{\.e}}\ \emph {et~al.}(2022)\citenamefont
  {B{\=u}tait{\.e}}, \citenamefont {Kupianskyi}, \citenamefont
  {{\v{C}}i{\v{z}}m{\'a}r},\ and\ \citenamefont {Phillips}}]{butaite2022build}%
  \BibitemOpen
  \bibfield  {author} {\bibinfo {author} {\bibfnamefont {Un{\.e}~G}\
  \bibnamefont {B{\=u}tait{\.e}}}, \bibinfo {author} {\bibfnamefont {Hlib}\
  \bibnamefont {Kupianskyi}}, \bibinfo {author} {\bibfnamefont
  {Tom{\'a}{\v{s}}}\ \bibnamefont {{\v{C}}i{\v{z}}m{\'a}r}}, \ and\ \bibinfo
  {author} {\bibfnamefont {David~B}\ \bibnamefont {Phillips}},\ }\bibfield
  {title} {\enquote {\bibinfo {title} {How to build the ``optical inverse'' of
  a multimode fibre},}\ }\href@noop {} {\bibfield  {journal} {\bibinfo
  {journal} {Intelligent Computing}\ }\textbf {\bibinfo {volume} {2022}}
  (\bibinfo {year} {2022})}\BibitemShut {NoStop}%
\bibitem [{\citenamefont {Molesky}\ \emph {et~al.}(2018)\citenamefont
  {Molesky}, \citenamefont {Lin}, \citenamefont {Piggott}, \citenamefont {Jin},
  \citenamefont {Vuckovi{\'c}},\ and\ \citenamefont
  {Rodriguez}}]{molesky2018inverse}%
  \BibitemOpen
  \bibfield  {author} {\bibinfo {author} {\bibfnamefont {Sean}\ \bibnamefont
  {Molesky}}, \bibinfo {author} {\bibfnamefont {Zin}\ \bibnamefont {Lin}},
  \bibinfo {author} {\bibfnamefont {Alexander~Y}\ \bibnamefont {Piggott}},
  \bibinfo {author} {\bibfnamefont {Weiliang}\ \bibnamefont {Jin}}, \bibinfo
  {author} {\bibfnamefont {Jelena}\ \bibnamefont {Vuckovi{\'c}}}, \ and\
  \bibinfo {author} {\bibfnamefont {Alejandro~W}\ \bibnamefont {Rodriguez}},\
  }\bibfield  {title} {\enquote {\bibinfo {title} {Inverse design in
  nanophotonics},}\ }\href@noop {} {\bibfield  {journal} {\bibinfo  {journal}
  {Nature Photonics}\ }\textbf {\bibinfo {volume} {12}},\ \bibinfo {pages}
  {659--670} (\bibinfo {year} {2018})}\BibitemShut {NoStop}%
\bibitem [{\citenamefont {Bogaerts}\ \emph {et~al.}(2020)\citenamefont
  {Bogaerts}, \citenamefont {P{\'e}rez}, \citenamefont {Capmany}, \citenamefont
  {Miller}, \citenamefont {Poon}, \citenamefont {Englund}, \citenamefont
  {Morichetti},\ and\ \citenamefont {Melloni}}]{bogaerts2020programmable}%
  \BibitemOpen
  \bibfield  {author} {\bibinfo {author} {\bibfnamefont {Wim}\ \bibnamefont
  {Bogaerts}}, \bibinfo {author} {\bibfnamefont {Daniel}\ \bibnamefont
  {P{\'e}rez}}, \bibinfo {author} {\bibfnamefont {Jos{\'e}}\ \bibnamefont
  {Capmany}}, \bibinfo {author} {\bibfnamefont {David~AB}\ \bibnamefont
  {Miller}}, \bibinfo {author} {\bibfnamefont {Joyce}\ \bibnamefont {Poon}},
  \bibinfo {author} {\bibfnamefont {Dirk}\ \bibnamefont {Englund}}, \bibinfo
  {author} {\bibfnamefont {Francesco}\ \bibnamefont {Morichetti}}, \ and\
  \bibinfo {author} {\bibfnamefont {Andrea}\ \bibnamefont {Melloni}},\
  }\bibfield  {title} {\enquote {\bibinfo {title} {Programmable photonic
  circuits},}\ }\href@noop {} {\bibfield  {journal} {\bibinfo  {journal}
  {Nature}\ }\textbf {\bibinfo {volume} {586}},\ \bibinfo {pages} {207--216}
  (\bibinfo {year} {2020})}\BibitemShut {NoStop}%
\bibitem [{\citenamefont {Resisi}\ \emph {et~al.}(2020)\citenamefont {Resisi},
  \citenamefont {Viernik}, \citenamefont {Popoff},\ and\ \citenamefont
  {Bromberg}}]{resisi2020wavefront}%
  \BibitemOpen
  \bibfield  {author} {\bibinfo {author} {\bibfnamefont {Shachar}\ \bibnamefont
  {Resisi}}, \bibinfo {author} {\bibfnamefont {Yehonatan}\ \bibnamefont
  {Viernik}}, \bibinfo {author} {\bibfnamefont {Sebastien~M}\ \bibnamefont
  {Popoff}}, \ and\ \bibinfo {author} {\bibfnamefont {Yaron}\ \bibnamefont
  {Bromberg}},\ }\bibfield  {title} {\enquote {\bibinfo {title} {Wavefront
  shaping in multimode fibers by transmission matrix engineering},}\
  }\href@noop {} {\bibfield  {journal} {\bibinfo  {journal} {APL Photonics}\
  }\textbf {\bibinfo {volume} {5}} (\bibinfo {year} {2020})}\BibitemShut
  {NoStop}%
\bibitem [{\citenamefont {Kupianskyi}\ \emph {et~al.}(2023)\citenamefont
  {Kupianskyi}, \citenamefont {Horsley},\ and\ \citenamefont
  {Phillips}}]{kupianskyi2023high}%
  \BibitemOpen
  \bibfield  {author} {\bibinfo {author} {\bibfnamefont {Hlib}\ \bibnamefont
  {Kupianskyi}}, \bibinfo {author} {\bibfnamefont {Simon~AR}\ \bibnamefont
  {Horsley}}, \ and\ \bibinfo {author} {\bibfnamefont {David~B}\ \bibnamefont
  {Phillips}},\ }\bibfield  {title} {\enquote {\bibinfo {title}
  {High-dimensional spatial mode sorting and optical circuit design using
  multi-plane light conversion},}\ }\href@noop {} {\bibfield  {journal}
  {\bibinfo  {journal} {APL Photonics}\ }\textbf {\bibinfo {volume} {8}}
  (\bibinfo {year} {2023})}\BibitemShut {NoStop}%
\bibitem [{\citenamefont {Labroille}\ \emph {et~al.}(2014)\citenamefont
  {Labroille}, \citenamefont {Denolle}, \citenamefont {Jian}, \citenamefont
  {Genevaux}, \citenamefont {Treps},\ and\ \citenamefont
  {Morizur}}]{labroille2014efficient}%
  \BibitemOpen
  \bibfield  {author} {\bibinfo {author} {\bibfnamefont {Guillaume}\
  \bibnamefont {Labroille}}, \bibinfo {author} {\bibfnamefont {Bertrand}\
  \bibnamefont {Denolle}}, \bibinfo {author} {\bibfnamefont {Pu}~\bibnamefont
  {Jian}}, \bibinfo {author} {\bibfnamefont {Philippe}\ \bibnamefont
  {Genevaux}}, \bibinfo {author} {\bibfnamefont {Nicolas}\ \bibnamefont
  {Treps}}, \ and\ \bibinfo {author} {\bibfnamefont {Jean-Fran{\c{c}}ois}\
  \bibnamefont {Morizur}},\ }\bibfield  {title} {\enquote {\bibinfo {title}
  {Efficient and mode selective spatial mode multiplexer based on multi-plane
  light conversion},}\ }\href@noop {} {\bibfield  {journal} {\bibinfo
  {journal} {Optics express}\ }\textbf {\bibinfo {volume} {22}},\ \bibinfo
  {pages} {15599--15607} (\bibinfo {year} {2014})}\BibitemShut {NoStop}%
\bibitem [{\citenamefont {Lin}\ \emph {et~al.}(2018)\citenamefont {Lin},
  \citenamefont {Rivenson}, \citenamefont {Yardimci}, \citenamefont {Veli},
  \citenamefont {Luo}, \citenamefont {Jarrahi},\ and\ \citenamefont
  {Ozcan}}]{lin2018all}%
  \BibitemOpen
  \bibfield  {author} {\bibinfo {author} {\bibfnamefont {Xing}\ \bibnamefont
  {Lin}}, \bibinfo {author} {\bibfnamefont {Yair}\ \bibnamefont {Rivenson}},
  \bibinfo {author} {\bibfnamefont {Nezih~T}\ \bibnamefont {Yardimci}},
  \bibinfo {author} {\bibfnamefont {Muhammed}\ \bibnamefont {Veli}}, \bibinfo
  {author} {\bibfnamefont {Yi}~\bibnamefont {Luo}}, \bibinfo {author}
  {\bibfnamefont {Mona}\ \bibnamefont {Jarrahi}}, \ and\ \bibinfo {author}
  {\bibfnamefont {Aydogan}\ \bibnamefont {Ozcan}},\ }\bibfield  {title}
  {\enquote {\bibinfo {title} {All-optical machine learning using diffractive
  deep neural networks},}\ }\href@noop {} {\bibfield  {journal} {\bibinfo
  {journal} {Science}\ }\textbf {\bibinfo {volume} {361}},\ \bibinfo {pages}
  {1004--1008} (\bibinfo {year} {2018})}\BibitemShut {NoStop}%
\bibitem [{\citenamefont {Pl{\"o}schner}\ \emph {et~al.}(2015)\citenamefont
  {Pl{\"o}schner}, \citenamefont {Tyc},\ and\ \citenamefont
  {{\v{C}}i{\v{z}}m{\'a}r}}]{ploschner2015seeing}%
  \BibitemOpen
  \bibfield  {author} {\bibinfo {author} {\bibfnamefont {Martin}\ \bibnamefont
  {Pl{\"o}schner}}, \bibinfo {author} {\bibfnamefont {Tom{\'a}{\v{s}}}\
  \bibnamefont {Tyc}}, \ and\ \bibinfo {author} {\bibfnamefont
  {Tom{\'a}{\v{s}}}\ \bibnamefont {{\v{C}}i{\v{z}}m{\'a}r}},\ }\bibfield
  {title} {\enquote {\bibinfo {title} {Seeing through chaos in multimode
  fibres},}\ }\href@noop {} {\bibfield  {journal} {\bibinfo  {journal} {Nature
  Photonics}\ }\textbf {\bibinfo {volume} {9}},\ \bibinfo {pages} {529--535}
  (\bibinfo {year} {2015})}\BibitemShut {NoStop}%
\bibitem [{\citenamefont {Morizur}\ \emph {et~al.}(2010)\citenamefont
  {Morizur}, \citenamefont {Nicholls}, \citenamefont {Jian}, \citenamefont
  {Armstrong}, \citenamefont {Treps}, \citenamefont {Hage}, \citenamefont
  {Hsu}, \citenamefont {Bowen}, \citenamefont {Janousek},\ and\ \citenamefont
  {Bachor}}]{morizur2010programmable}%
  \BibitemOpen
  \bibfield  {author} {\bibinfo {author} {\bibfnamefont {Jean-Fran{\c{c}}ois}\
  \bibnamefont {Morizur}}, \bibinfo {author} {\bibfnamefont {Lachlan}\
  \bibnamefont {Nicholls}}, \bibinfo {author} {\bibfnamefont {Pu}~\bibnamefont
  {Jian}}, \bibinfo {author} {\bibfnamefont {Seiji}\ \bibnamefont {Armstrong}},
  \bibinfo {author} {\bibfnamefont {Nicolas}\ \bibnamefont {Treps}}, \bibinfo
  {author} {\bibfnamefont {Boris}\ \bibnamefont {Hage}}, \bibinfo {author}
  {\bibfnamefont {Magnus}\ \bibnamefont {Hsu}}, \bibinfo {author}
  {\bibfnamefont {Warwick}\ \bibnamefont {Bowen}}, \bibinfo {author}
  {\bibfnamefont {Jiri}\ \bibnamefont {Janousek}}, \ and\ \bibinfo {author}
  {\bibfnamefont {Hans-A}\ \bibnamefont {Bachor}},\ }\bibfield  {title}
  {\enquote {\bibinfo {title} {Programmable unitary spatial mode
  manipulation},}\ }\href@noop {} {\bibfield  {journal} {\bibinfo  {journal}
  {JOSA A}\ }\textbf {\bibinfo {volume} {27}},\ \bibinfo {pages} {2524--2531}
  (\bibinfo {year} {2010})}\BibitemShut {NoStop}%
\bibitem [{\citenamefont {Wang}\ and\ \citenamefont
  {Piestun}(2018)}]{wang2018dynamic}%
  \BibitemOpen
  \bibfield  {author} {\bibinfo {author} {\bibfnamefont {Haiyan}\ \bibnamefont
  {Wang}}\ and\ \bibinfo {author} {\bibfnamefont {Rafael}\ \bibnamefont
  {Piestun}},\ }\bibfield  {title} {\enquote {\bibinfo {title} {Dynamic 2d
  implementation of 3d diffractive optics},}\ }\href@noop {} {\bibfield
  {journal} {\bibinfo  {journal} {Optica}\ }\textbf {\bibinfo {volume} {5}},\
  \bibinfo {pages} {1220--1228} (\bibinfo {year} {2018})}\BibitemShut {NoStop}%
\bibitem [{\citenamefont {Fontaine}\ \emph {et~al.}(2019)\citenamefont
  {Fontaine}, \citenamefont {Ryf}, \citenamefont {Chen}, \citenamefont
  {Neilson}, \citenamefont {Kim},\ and\ \citenamefont
  {Carpenter}}]{fontaine2019laguerre}%
  \BibitemOpen
  \bibfield  {author} {\bibinfo {author} {\bibfnamefont {Nicolas~K}\
  \bibnamefont {Fontaine}}, \bibinfo {author} {\bibfnamefont {Roland}\
  \bibnamefont {Ryf}}, \bibinfo {author} {\bibfnamefont {Haoshuo}\ \bibnamefont
  {Chen}}, \bibinfo {author} {\bibfnamefont {David~T}\ \bibnamefont {Neilson}},
  \bibinfo {author} {\bibfnamefont {Kwangwoong}\ \bibnamefont {Kim}}, \ and\
  \bibinfo {author} {\bibfnamefont {Joel}\ \bibnamefont {Carpenter}},\
  }\bibfield  {title} {\enquote {\bibinfo {title} {Laguerre-gaussian mode
  sorter},}\ }\href@noop {} {\bibfield  {journal} {\bibinfo  {journal} {Nature
  communications}\ }\textbf {\bibinfo {volume} {10}},\ \bibinfo {pages} {1--7}
  (\bibinfo {year} {2019})}\BibitemShut {NoStop}%
\bibitem [{\citenamefont {Brandt}\ \emph {et~al.}(2020)\citenamefont {Brandt},
  \citenamefont {Hiekkam{\"a}ki}, \citenamefont {Bouchard}, \citenamefont
  {Huber},\ and\ \citenamefont {Fickler}}]{brandt2020high}%
  \BibitemOpen
  \bibfield  {author} {\bibinfo {author} {\bibfnamefont {Florian}\ \bibnamefont
  {Brandt}}, \bibinfo {author} {\bibfnamefont {Markus}\ \bibnamefont
  {Hiekkam{\"a}ki}}, \bibinfo {author} {\bibfnamefont {Fr{\'e}d{\'e}ric}\
  \bibnamefont {Bouchard}}, \bibinfo {author} {\bibfnamefont {Marcus}\
  \bibnamefont {Huber}}, \ and\ \bibinfo {author} {\bibfnamefont {Robert}\
  \bibnamefont {Fickler}},\ }\bibfield  {title} {\enquote {\bibinfo {title}
  {High-dimensional quantum gates using full-field spatial modes of photons},}\
  }\href@noop {} {\bibfield  {journal} {\bibinfo  {journal} {Optica}\ }\textbf
  {\bibinfo {volume} {7}},\ \bibinfo {pages} {98--107} (\bibinfo {year}
  {2020})}\BibitemShut {NoStop}%
\bibitem [{\citenamefont {Fickler}\ \emph {et~al.}(2020)\citenamefont
  {Fickler}, \citenamefont {Bouchard}, \citenamefont {Giese}, \citenamefont
  {Grillo}, \citenamefont {Leuchs},\ and\ \citenamefont
  {Karimi}}]{fickler2020full}%
  \BibitemOpen
  \bibfield  {author} {\bibinfo {author} {\bibfnamefont {Robert}\ \bibnamefont
  {Fickler}}, \bibinfo {author} {\bibfnamefont {Fr{\'e}d{\'e}ric}\ \bibnamefont
  {Bouchard}}, \bibinfo {author} {\bibfnamefont {Enno}\ \bibnamefont {Giese}},
  \bibinfo {author} {\bibfnamefont {Vincenzo}\ \bibnamefont {Grillo}}, \bibinfo
  {author} {\bibfnamefont {Gerd}\ \bibnamefont {Leuchs}}, \ and\ \bibinfo
  {author} {\bibfnamefont {Ebrahim}\ \bibnamefont {Karimi}},\ }\bibfield
  {title} {\enquote {\bibinfo {title} {Full-field mode sorter using two
  optimized phase transformations for high-dimensional quantum cryptography},}\
  }\href@noop {} {\bibfield  {journal} {\bibinfo  {journal} {Journal of
  Optics}\ }\textbf {\bibinfo {volume} {22}},\ \bibinfo {pages} {024001}
  (\bibinfo {year} {2020})}\BibitemShut {NoStop}%
\bibitem [{\citenamefont {Mounaix}\ \emph {et~al.}(2020)\citenamefont
  {Mounaix}, \citenamefont {Fontaine}, \citenamefont {Neilson}, \citenamefont
  {Ryf}, \citenamefont {Chen}, \citenamefont {Alvarado-Zacarias},\ and\
  \citenamefont {Carpenter}}]{mounaix2020time}%
  \BibitemOpen
  \bibfield  {author} {\bibinfo {author} {\bibfnamefont {Mickael}\ \bibnamefont
  {Mounaix}}, \bibinfo {author} {\bibfnamefont {Nicolas~K}\ \bibnamefont
  {Fontaine}}, \bibinfo {author} {\bibfnamefont {David~T}\ \bibnamefont
  {Neilson}}, \bibinfo {author} {\bibfnamefont {Roland}\ \bibnamefont {Ryf}},
  \bibinfo {author} {\bibfnamefont {Haoshuo}\ \bibnamefont {Chen}}, \bibinfo
  {author} {\bibfnamefont {Juan~Carlos}\ \bibnamefont {Alvarado-Zacarias}}, \
  and\ \bibinfo {author} {\bibfnamefont {Joel}\ \bibnamefont {Carpenter}},\
  }\bibfield  {title} {\enquote {\bibinfo {title} {Time reversed optical waves
  by arbitrary vector spatiotemporal field generation},}\ }\href@noop {}
  {\bibfield  {journal} {\bibinfo  {journal} {Nature communications}\ }\textbf
  {\bibinfo {volume} {11}},\ \bibinfo {pages} {1--7} (\bibinfo {year}
  {2020})}\BibitemShut {NoStop}%
\bibitem [{\citenamefont {Lib}\ \emph {et~al.}(2021)\citenamefont {Lib},
  \citenamefont {Sulimany},\ and\ \citenamefont
  {Bromberg}}]{lib2021reconfigurable}%
  \BibitemOpen
  \bibfield  {author} {\bibinfo {author} {\bibfnamefont {Ohad}\ \bibnamefont
  {Lib}}, \bibinfo {author} {\bibfnamefont {Kfir}\ \bibnamefont {Sulimany}}, \
  and\ \bibinfo {author} {\bibfnamefont {Yaron}\ \bibnamefont {Bromberg}},\
  }\bibfield  {title} {\enquote {\bibinfo {title} {Reconfigurable synthesizer
  for quantum information processing of high-dimensional entangled photons},}\
  }\href@noop {} {\bibfield  {journal} {\bibinfo  {journal} {arXiv preprint
  arXiv:2108.02258}\ } (\bibinfo {year} {2021})}\BibitemShut {NoStop}%
\bibitem [{\citenamefont {Korichi}\ \emph {et~al.}(2023)\citenamefont
  {Korichi}, \citenamefont {Hiekkam{\"a}ki},\ and\ \citenamefont
  {Fickler}}]{korichi2023high}%
  \BibitemOpen
  \bibfield  {author} {\bibinfo {author} {\bibfnamefont {Oussama}\ \bibnamefont
  {Korichi}}, \bibinfo {author} {\bibfnamefont {Markus}\ \bibnamefont
  {Hiekkam{\"a}ki}}, \ and\ \bibinfo {author} {\bibfnamefont {Robert}\
  \bibnamefont {Fickler}},\ }\bibfield  {title} {\enquote {\bibinfo {title}
  {High-efficiency interface between multi-mode and single-mode fibers},}\
  }\href@noop {} {\bibfield  {journal} {\bibinfo  {journal} {Optics Letters}\
  }\textbf {\bibinfo {volume} {48}},\ \bibinfo {pages} {1000--1003} (\bibinfo
  {year} {2023})}\BibitemShut {NoStop}%
\bibitem [{\citenamefont {Hashimoto}\ \emph {et~al.}(2005)\citenamefont
  {Hashimoto}, \citenamefont {Saida}, \citenamefont {Ogawa}, \citenamefont
  {Kohtoku}, \citenamefont {Shibata},\ and\ \citenamefont
  {Takahashi}}]{hashimoto2005optical}%
  \BibitemOpen
  \bibfield  {author} {\bibinfo {author} {\bibfnamefont {T}~\bibnamefont
  {Hashimoto}}, \bibinfo {author} {\bibfnamefont {T}~\bibnamefont {Saida}},
  \bibinfo {author} {\bibfnamefont {I}~\bibnamefont {Ogawa}}, \bibinfo {author}
  {\bibfnamefont {M}~\bibnamefont {Kohtoku}}, \bibinfo {author} {\bibfnamefont
  {Tomohiro}\ \bibnamefont {Shibata}}, \ and\ \bibinfo {author} {\bibfnamefont
  {Hiroshi}\ \bibnamefont {Takahashi}},\ }\bibfield  {title} {\enquote
  {\bibinfo {title} {Optical circuit design based on a wavefront-matching
  method},}\ }\href@noop {} {\bibfield  {journal} {\bibinfo  {journal} {Optics
  letters}\ }\textbf {\bibinfo {volume} {30}},\ \bibinfo {pages} {2620--2622}
  (\bibinfo {year} {2005})}\BibitemShut {NoStop}%
\bibitem [{\citenamefont {Barr{\'e}}\ and\ \citenamefont
  {Jesacher}(2022)}]{barre2022inverse}%
  \BibitemOpen
  \bibfield  {author} {\bibinfo {author} {\bibfnamefont {Nicolas}\ \bibnamefont
  {Barr{\'e}}}\ and\ \bibinfo {author} {\bibfnamefont {Alexander}\ \bibnamefont
  {Jesacher}},\ }\bibfield  {title} {\enquote {\bibinfo {title} {Inverse design
  of gradient-index volume multimode converters},}\ }\href@noop {} {\bibfield
  {journal} {\bibinfo  {journal} {Optics Express}\ }\textbf {\bibinfo {volume}
  {30}},\ \bibinfo {pages} {10573--10587} (\bibinfo {year} {2022})}\BibitemShut
  {NoStop}%
\bibitem [{\citenamefont {Lee}(1979)}]{lee1979binary}%
  \BibitemOpen
  \bibfield  {author} {\bibinfo {author} {\bibfnamefont {Wai-Hon}\ \bibnamefont
  {Lee}},\ }\bibfield  {title} {\enquote {\bibinfo {title} {Binary
  computer-generated holograms},}\ }\href@noop {} {\bibfield  {journal}
  {\bibinfo  {journal} {Applied Optics}\ }\textbf {\bibinfo {volume} {18}},\
  \bibinfo {pages} {3661--3669} (\bibinfo {year} {1979})}\BibitemShut {NoStop}%
\bibitem [{\citenamefont {Mitchell}\ \emph {et~al.}(2016)\citenamefont
  {Mitchell}, \citenamefont {Turtaev}, \citenamefont {Padgett}, \citenamefont
  {{\v{C}}i{\v{z}}m{\'a}r},\ and\ \citenamefont {Phillips}}]{mitchell2016high}%
  \BibitemOpen
  \bibfield  {author} {\bibinfo {author} {\bibfnamefont {Kevin~J}\ \bibnamefont
  {Mitchell}}, \bibinfo {author} {\bibfnamefont {Sergey}\ \bibnamefont
  {Turtaev}}, \bibinfo {author} {\bibfnamefont {Miles~J}\ \bibnamefont
  {Padgett}}, \bibinfo {author} {\bibfnamefont {Tom{\'a}{\v{s}}}\ \bibnamefont
  {{\v{C}}i{\v{z}}m{\'a}r}}, \ and\ \bibinfo {author} {\bibfnamefont {David~B}\
  \bibnamefont {Phillips}},\ }\bibfield  {title} {\enquote {\bibinfo {title}
  {High-speed spatial control of the intensity, phase and polarisation of
  vector beams using a digital micro-mirror device},}\ }\href@noop {}
  {\bibfield  {journal} {\bibinfo  {journal} {Optics express}\ }\textbf
  {\bibinfo {volume} {24}},\ \bibinfo {pages} {29269--29282} (\bibinfo {year}
  {2016})}\BibitemShut {NoStop}%
\bibitem [{\citenamefont {Fontaine}\ \emph {et~al.}(2017)\citenamefont
  {Fontaine}, \citenamefont {Chen}, \citenamefont {Ryf}, \citenamefont
  {Neilson}, \citenamefont {Alvarado}, \citenamefont {Van~Weerdenburg},
  \citenamefont {Amezcua-Correa}, \citenamefont {Okonkwo},\ and\ \citenamefont
  {Carpenter}}]{fontaine2017programmable}%
  \BibitemOpen
  \bibfield  {author} {\bibinfo {author} {\bibfnamefont {Nicolas~K}\
  \bibnamefont {Fontaine}}, \bibinfo {author} {\bibfnamefont {Haoshuo}\
  \bibnamefont {Chen}}, \bibinfo {author} {\bibfnamefont {Roland}\ \bibnamefont
  {Ryf}}, \bibinfo {author} {\bibfnamefont {David}\ \bibnamefont {Neilson}},
  \bibinfo {author} {\bibfnamefont {Juan~Carlos}\ \bibnamefont {Alvarado}},
  \bibinfo {author} {\bibfnamefont {John}\ \bibnamefont {Van~Weerdenburg}},
  \bibinfo {author} {\bibfnamefont {Rodrigo}\ \bibnamefont {Amezcua-Correa}},
  \bibinfo {author} {\bibfnamefont {Chigo}\ \bibnamefont {Okonkwo}}, \ and\
  \bibinfo {author} {\bibfnamefont {Joel}\ \bibnamefont {Carpenter}},\
  }\bibfield  {title} {\enquote {\bibinfo {title} {Programmable vector mode
  multiplexer},}\ }in\ \href@noop {} {\emph {\bibinfo {booktitle} {2017
  European Conference on Optical Communication (ECOC)}}}\ (\bibinfo
  {organization} {IEEE},\ \bibinfo {year} {2017})\ pp.\ \bibinfo {pages}
  {1--3}\BibitemShut {NoStop}%
\bibitem [{\citenamefont {Li}\ \emph {et~al.}(2021{\natexlab{a}})\citenamefont
  {Li}, \citenamefont {Saunders}, \citenamefont {Lum}, \citenamefont
  {Murray-Bruce}, \citenamefont {Goyal}, \citenamefont
  {{\v{C}}i{\v{z}}m{\'a}r},\ and\ \citenamefont
  {Phillips}}]{li2021compressively}%
  \BibitemOpen
  \bibfield  {author} {\bibinfo {author} {\bibfnamefont {Shuhui}\ \bibnamefont
  {Li}}, \bibinfo {author} {\bibfnamefont {Charles}\ \bibnamefont {Saunders}},
  \bibinfo {author} {\bibfnamefont {Daniel~J}\ \bibnamefont {Lum}}, \bibinfo
  {author} {\bibfnamefont {John}\ \bibnamefont {Murray-Bruce}}, \bibinfo
  {author} {\bibfnamefont {Vivek~K}\ \bibnamefont {Goyal}}, \bibinfo {author}
  {\bibfnamefont {Tom{\'a}{\v{s}}}\ \bibnamefont {{\v{C}}i{\v{z}}m{\'a}r}}, \
  and\ \bibinfo {author} {\bibfnamefont {David~B}\ \bibnamefont {Phillips}},\
  }\bibfield  {title} {\enquote {\bibinfo {title} {Compressively sampling the
  optical transmission matrix of a multimode fibre},}\ }\href@noop {}
  {\bibfield  {journal} {\bibinfo  {journal} {Light: Science \& Applications}\
  }\textbf {\bibinfo {volume} {10}},\ \bibinfo {pages} {1--15} (\bibinfo {year}
  {2021}{\natexlab{a}})}\BibitemShut {NoStop}%
\bibitem [{\citenamefont {Gu}\ \emph {et~al.}(2015)\citenamefont {Gu},
  \citenamefont {Mahalati},\ and\ \citenamefont {Kahn}}]{gu2015design}%
  \BibitemOpen
  \bibfield  {author} {\bibinfo {author} {\bibfnamefont {Ruo~Yu}\ \bibnamefont
  {Gu}}, \bibinfo {author} {\bibfnamefont {Reza~Nasiri}\ \bibnamefont
  {Mahalati}}, \ and\ \bibinfo {author} {\bibfnamefont {Joseph~M}\ \bibnamefont
  {Kahn}},\ }\bibfield  {title} {\enquote {\bibinfo {title} {Design of flexible
  multi-mode fiber endoscope},}\ }\href@noop {} {\bibfield  {journal} {\bibinfo
   {journal} {Optics express}\ }\textbf {\bibinfo {volume} {23}},\ \bibinfo
  {pages} {26905--26918} (\bibinfo {year} {2015})}\BibitemShut {NoStop}%
\bibitem [{\citenamefont {Gordon}\ \emph {et~al.}(2019)\citenamefont {Gordon},
  \citenamefont {Gataric}, \citenamefont {Ramos}, \citenamefont {Mouthaan},
  \citenamefont {Williams}, \citenamefont {Yoon}, \citenamefont {Wilkinson},\
  and\ \citenamefont {Bohndiek}}]{gordon2019characterizing}%
  \BibitemOpen
  \bibfield  {author} {\bibinfo {author} {\bibfnamefont {George~SD}\
  \bibnamefont {Gordon}}, \bibinfo {author} {\bibfnamefont {Milana}\
  \bibnamefont {Gataric}}, \bibinfo {author} {\bibfnamefont {Alberto Gil~CP}\
  \bibnamefont {Ramos}}, \bibinfo {author} {\bibfnamefont {Ralf}\ \bibnamefont
  {Mouthaan}}, \bibinfo {author} {\bibfnamefont {Calum}\ \bibnamefont
  {Williams}}, \bibinfo {author} {\bibfnamefont {Jonghee}\ \bibnamefont
  {Yoon}}, \bibinfo {author} {\bibfnamefont {Timothy~D}\ \bibnamefont
  {Wilkinson}}, \ and\ \bibinfo {author} {\bibfnamefont {Sarah~E}\ \bibnamefont
  {Bohndiek}},\ }\bibfield  {title} {\enquote {\bibinfo {title} {Characterizing
  optical fiber transmission matrices using metasurface reflector stacks for
  lensless imaging without distal access},}\ }\href@noop {} {\bibfield
  {journal} {\bibinfo  {journal} {Physical Review X}\ }\textbf {\bibinfo
  {volume} {9}},\ \bibinfo {pages} {041050} (\bibinfo {year}
  {2019})}\BibitemShut {NoStop}%
\bibitem [{\citenamefont {Li}\ \emph {et~al.}(2021{\natexlab{b}})\citenamefont
  {Li}, \citenamefont {Horsley}, \citenamefont {Tyc}, \citenamefont
  {{\v{C}}i{\v{z}}m{\'a}r},\ and\ \citenamefont {Phillips}}]{li2021memory}%
  \BibitemOpen
  \bibfield  {author} {\bibinfo {author} {\bibfnamefont {Shuhui}\ \bibnamefont
  {Li}}, \bibinfo {author} {\bibfnamefont {Simon~AR}\ \bibnamefont {Horsley}},
  \bibinfo {author} {\bibfnamefont {Tom{\'a}{\v{s}}}\ \bibnamefont {Tyc}},
  \bibinfo {author} {\bibfnamefont {Tom{\'a}{\v{s}}}\ \bibnamefont
  {{\v{C}}i{\v{z}}m{\'a}r}}, \ and\ \bibinfo {author} {\bibfnamefont {David~B}\
  \bibnamefont {Phillips}},\ }\bibfield  {title} {\enquote {\bibinfo {title}
  {Memory effect assisted imaging through multimode optical fibres},}\
  }\href@noop {} {\bibfield  {journal} {\bibinfo  {journal} {Nature
  Communications}\ }\textbf {\bibinfo {volume} {12}},\ \bibinfo {pages} {1--13}
  (\bibinfo {year} {2021}{\natexlab{b}})}\BibitemShut {NoStop}%
\bibitem [{\citenamefont {Luo}\ \emph {et~al.}(2022)\citenamefont {Luo},
  \citenamefont {Zhao}, \citenamefont {Li}, \citenamefont
  {{\c{C}}etinta{\c{s}}}, \citenamefont {Rivenson}, \citenamefont {Jarrahi},\
  and\ \citenamefont {Ozcan}}]{luo2022computational}%
  \BibitemOpen
  \bibfield  {author} {\bibinfo {author} {\bibfnamefont {Yi}~\bibnamefont
  {Luo}}, \bibinfo {author} {\bibfnamefont {Yifan}\ \bibnamefont {Zhao}},
  \bibinfo {author} {\bibfnamefont {Jingxi}\ \bibnamefont {Li}}, \bibinfo
  {author} {\bibfnamefont {Ege}\ \bibnamefont {{\c{C}}etinta{\c{s}}}}, \bibinfo
  {author} {\bibfnamefont {Yair}\ \bibnamefont {Rivenson}}, \bibinfo {author}
  {\bibfnamefont {Mona}\ \bibnamefont {Jarrahi}}, \ and\ \bibinfo {author}
  {\bibfnamefont {Aydogan}\ \bibnamefont {Ozcan}},\ }\bibfield  {title}
  {\enquote {\bibinfo {title} {Computational imaging without a computer: seeing
  through random diffusers at the speed of light},}\ }\href@noop {} {\bibfield
  {journal} {\bibinfo  {journal} {eLight}\ }\textbf {\bibinfo {volume} {2}},\
  \bibinfo {pages} {4} (\bibinfo {year} {2022})}\BibitemShut {NoStop}%
\bibitem [{\citenamefont {Mengu}\ and\ \citenamefont
  {Ozcan}(2022)}]{mengu2022all}%
  \BibitemOpen
  \bibfield  {author} {\bibinfo {author} {\bibfnamefont {Deniz}\ \bibnamefont
  {Mengu}}\ and\ \bibinfo {author} {\bibfnamefont {Aydogan}\ \bibnamefont
  {Ozcan}},\ }\bibfield  {title} {\enquote {\bibinfo {title} {All-optical phase
  recovery: diffractive computing for quantitative phase imaging},}\
  }\href@noop {} {\bibfield  {journal} {\bibinfo  {journal} {Advanced Optical
  Materials}\ }\textbf {\bibinfo {volume} {10}},\ \bibinfo {pages} {2200281}
  (\bibinfo {year} {2022})}\BibitemShut {NoStop}%
\bibitem [{\citenamefont {Borhani}\ \emph {et~al.}(2018)\citenamefont
  {Borhani}, \citenamefont {Kakkava}, \citenamefont {Moser},\ and\
  \citenamefont {Psaltis}}]{borhani2018learning}%
  \BibitemOpen
  \bibfield  {author} {\bibinfo {author} {\bibfnamefont {Navid}\ \bibnamefont
  {Borhani}}, \bibinfo {author} {\bibfnamefont {Eirini}\ \bibnamefont
  {Kakkava}}, \bibinfo {author} {\bibfnamefont {Christophe}\ \bibnamefont
  {Moser}}, \ and\ \bibinfo {author} {\bibfnamefont {Demetri}\ \bibnamefont
  {Psaltis}},\ }\bibfield  {title} {\enquote {\bibinfo {title} {Learning to see
  through multimode fibers},}\ }\href@noop {} {\bibfield  {journal} {\bibinfo
  {journal} {Optica}\ }\textbf {\bibinfo {volume} {5}},\ \bibinfo {pages}
  {960--966} (\bibinfo {year} {2018})}\BibitemShut {NoStop}%
\bibitem [{\citenamefont {Caramazza}\ \emph {et~al.}(2019)\citenamefont
  {Caramazza}, \citenamefont {Moran}, \citenamefont {Murray-Smith},\ and\
  \citenamefont {Faccio}}]{caramazza2019transmission}%
  \BibitemOpen
  \bibfield  {author} {\bibinfo {author} {\bibfnamefont {Piergiorgio}\
  \bibnamefont {Caramazza}}, \bibinfo {author} {\bibfnamefont {Ois{\'\i}n}\
  \bibnamefont {Moran}}, \bibinfo {author} {\bibfnamefont {Roderick}\
  \bibnamefont {Murray-Smith}}, \ and\ \bibinfo {author} {\bibfnamefont
  {Daniele}\ \bibnamefont {Faccio}},\ }\bibfield  {title} {\enquote {\bibinfo
  {title} {Transmission of natural scene images through a multimode fibre},}\
  }\href@noop {} {\bibfield  {journal} {\bibinfo  {journal} {Nature
  communications}\ }\textbf {\bibinfo {volume} {10}},\ \bibinfo {pages} {2029}
  (\bibinfo {year} {2019})}\BibitemShut {NoStop}%
\bibitem [{\citenamefont {Resisi}\ \emph {et~al.}(2021)\citenamefont {Resisi},
  \citenamefont {Popoff},\ and\ \citenamefont {Bromberg}}]{resisi2021image}%
  \BibitemOpen
  \bibfield  {author} {\bibinfo {author} {\bibfnamefont {Shachar}\ \bibnamefont
  {Resisi}}, \bibinfo {author} {\bibfnamefont {Sebastien~M}\ \bibnamefont
  {Popoff}}, \ and\ \bibinfo {author} {\bibfnamefont {Yaron}\ \bibnamefont
  {Bromberg}},\ }\bibfield  {title} {\enquote {\bibinfo {title} {Image
  transmission through a dynamically perturbed multimode fiber by deep
  learning},}\ }\href@noop {} {\bibfield  {journal} {\bibinfo  {journal} {Laser
  \& Photonics Reviews}\ }\textbf {\bibinfo {volume} {15}},\ \bibinfo {pages}
  {2000553} (\bibinfo {year} {2021})}\BibitemShut {NoStop}%
\bibitem [{\citenamefont {Abdulaziz}\ \emph {et~al.}(2022)\citenamefont
  {Abdulaziz}, \citenamefont {Mekhail}, \citenamefont {Altmann}, \citenamefont
  {Padgett},\ and\ \citenamefont {McLaughlin}}]{abdulaziz2022robust}%
  \BibitemOpen
  \bibfield  {author} {\bibinfo {author} {\bibfnamefont {Abdullah}\
  \bibnamefont {Abdulaziz}}, \bibinfo {author} {\bibfnamefont {Simon~Peter}\
  \bibnamefont {Mekhail}}, \bibinfo {author} {\bibfnamefont {Yoann}\
  \bibnamefont {Altmann}}, \bibinfo {author} {\bibfnamefont {Miles~J}\
  \bibnamefont {Padgett}}, \ and\ \bibinfo {author} {\bibfnamefont {Stephen}\
  \bibnamefont {McLaughlin}},\ }\bibfield  {title} {\enquote {\bibinfo {title}
  {Robust real-time imaging through flexible multimode fibers},}\ }\href@noop
  {} {\bibfield  {journal} {\bibinfo  {journal} {arXiv preprint
  arXiv:2210.13883}\ } (\bibinfo {year} {2022})}\BibitemShut {NoStop}%
\bibitem [{\citenamefont {Kim}\ and\ \citenamefont
  {Menon}(2014)}]{kim2014ultra}%
  \BibitemOpen
  \bibfield  {author} {\bibinfo {author} {\bibfnamefont {Ganghun}\ \bibnamefont
  {Kim}}\ and\ \bibinfo {author} {\bibfnamefont {Rajesh}\ \bibnamefont
  {Menon}},\ }\bibfield  {title} {\enquote {\bibinfo {title} {An ultra-small
  three dimensional computational microscope},}\ }\href@noop {} {\bibfield
  {journal} {\bibinfo  {journal} {Applied Physics Letters}\ }\textbf {\bibinfo
  {volume} {105}},\ \bibinfo {pages} {061114} (\bibinfo {year}
  {2014})}\BibitemShut {NoStop}%
\bibitem [{\citenamefont {Guo}\ \emph {et~al.}(2022)\citenamefont {Guo},
  \citenamefont {Nelson}, \citenamefont {Regier}, \citenamefont {Davis},
  \citenamefont {Jorgensen}, \citenamefont {Shepherd},\ and\ \citenamefont
  {Menon}}]{guo2022scan}%
  \BibitemOpen
  \bibfield  {author} {\bibinfo {author} {\bibfnamefont {Ruipeng}\ \bibnamefont
  {Guo}}, \bibinfo {author} {\bibfnamefont {Soren}\ \bibnamefont {Nelson}},
  \bibinfo {author} {\bibfnamefont {Matthew}\ \bibnamefont {Regier}}, \bibinfo
  {author} {\bibfnamefont {M~Wayne}\ \bibnamefont {Davis}}, \bibinfo {author}
  {\bibfnamefont {Erik~M}\ \bibnamefont {Jorgensen}}, \bibinfo {author}
  {\bibfnamefont {Jason}\ \bibnamefont {Shepherd}}, \ and\ \bibinfo {author}
  {\bibfnamefont {Rajesh}\ \bibnamefont {Menon}},\ }\bibfield  {title}
  {\enquote {\bibinfo {title} {Scan-less machine-learning-enabled incoherent
  microscopy for minimally-invasive deep-brain imaging},}\ }\href@noop {}
  {\bibfield  {journal} {\bibinfo  {journal} {Optics Express}\ }\textbf
  {\bibinfo {volume} {30}},\ \bibinfo {pages} {1546--1554} (\bibinfo {year}
  {2022})}\BibitemShut {NoStop}%
\bibitem [{\citenamefont {Pastor}\ \emph {et~al.}(2021)\citenamefont {Pastor},
  \citenamefont {Lundeen},\ and\ \citenamefont
  {Marquardt}}]{pastor2021arbitrary}%
  \BibitemOpen
  \bibfield  {author} {\bibinfo {author} {\bibfnamefont
  {V{\'\i}ctor~L{\'o}pez}\ \bibnamefont {Pastor}}, \bibinfo {author}
  {\bibfnamefont {Jeff}\ \bibnamefont {Lundeen}}, \ and\ \bibinfo {author}
  {\bibfnamefont {Florian}\ \bibnamefont {Marquardt}},\ }\bibfield  {title}
  {\enquote {\bibinfo {title} {Arbitrary optical wave evolution with fourier
  transforms and phase masks},}\ }\href@noop {} {\bibfield  {journal} {\bibinfo
   {journal} {Optics Express}\ }\textbf {\bibinfo {volume} {29}},\ \bibinfo
  {pages} {38441--38450} (\bibinfo {year} {2021})}\BibitemShut {NoStop}%
\bibitem [{\citenamefont {Berkhout}\ \emph {et~al.}(2010)\citenamefont
  {Berkhout}, \citenamefont {Lavery}, \citenamefont {Courtial}, \citenamefont
  {Beijersbergen},\ and\ \citenamefont {Padgett}}]{berkhout2010efficient}%
  \BibitemOpen
  \bibfield  {author} {\bibinfo {author} {\bibfnamefont {Gregorius~CG}\
  \bibnamefont {Berkhout}}, \bibinfo {author} {\bibfnamefont {Martin~PJ}\
  \bibnamefont {Lavery}}, \bibinfo {author} {\bibfnamefont {Johannes}\
  \bibnamefont {Courtial}}, \bibinfo {author} {\bibfnamefont {Marco~W}\
  \bibnamefont {Beijersbergen}}, \ and\ \bibinfo {author} {\bibfnamefont
  {Miles~J}\ \bibnamefont {Padgett}},\ }\bibfield  {title} {\enquote {\bibinfo
  {title} {Efficient sorting of orbital angular momentum states of light},}\
  }\href@noop {} {\bibfield  {journal} {\bibinfo  {journal} {Physical review
  letters}\ }\textbf {\bibinfo {volume} {105}},\ \bibinfo {pages} {153601}
  (\bibinfo {year} {2010})}\BibitemShut {NoStop}%
\bibitem [{\citenamefont {Gover}\ \emph {et~al.}(1976)\citenamefont {Gover},
  \citenamefont {Lee},\ and\ \citenamefont {Yariv}}]{gover1976direct}%
  \BibitemOpen
  \bibfield  {author} {\bibinfo {author} {\bibfnamefont {A}~\bibnamefont
  {Gover}}, \bibinfo {author} {\bibfnamefont {CP}~\bibnamefont {Lee}}, \ and\
  \bibinfo {author} {\bibfnamefont {A}~\bibnamefont {Yariv}},\ }\bibfield
  {title} {\enquote {\bibinfo {title} {Direct transmission of pictorial
  information in multimode optical fibers},}\ }\href@noop {} {\bibfield
  {journal} {\bibinfo  {journal} {JOSA}\ }\textbf {\bibinfo {volume} {66}},\
  \bibinfo {pages} {306--311} (\bibinfo {year} {1976})}\BibitemShut {NoStop}%
\bibitem [{\citenamefont {Badt}\ and\ \citenamefont
  {Katz}(2022)}]{badt2022real}%
  \BibitemOpen
  \bibfield  {author} {\bibinfo {author} {\bibfnamefont {Noam}\ \bibnamefont
  {Badt}}\ and\ \bibinfo {author} {\bibfnamefont {Ori}\ \bibnamefont {Katz}},\
  }\bibfield  {title} {\enquote {\bibinfo {title} {Real-time holographic
  lensless micro-endoscopy through flexible fibers via fiber bundle distal
  holography},}\ }\href@noop {} {\bibfield  {journal} {\bibinfo  {journal}
  {Nature Communications}\ }\textbf {\bibinfo {volume} {13}},\ \bibinfo {pages}
  {6055} (\bibinfo {year} {2022})}\BibitemShut {NoStop}%
\bibitem [{\citenamefont {Choudhury}\ \emph {et~al.}(2020)\citenamefont
  {Choudhury}, \citenamefont {McNicholl}, \citenamefont {Repetti},
  \citenamefont {Gris-S{\'a}nchez}, \citenamefont {Li}, \citenamefont
  {Phillips}, \citenamefont {Whyte}, \citenamefont {Birks}, \citenamefont
  {Wiaux},\ and\ \citenamefont {Thomson}}]{choudhury2020computational}%
  \BibitemOpen
  \bibfield  {author} {\bibinfo {author} {\bibfnamefont {Debaditya}\
  \bibnamefont {Choudhury}}, \bibinfo {author} {\bibfnamefont {Duncan~K}\
  \bibnamefont {McNicholl}}, \bibinfo {author} {\bibfnamefont {Audrey}\
  \bibnamefont {Repetti}}, \bibinfo {author} {\bibfnamefont {Itandehui}\
  \bibnamefont {Gris-S{\'a}nchez}}, \bibinfo {author} {\bibfnamefont {Shuhui}\
  \bibnamefont {Li}}, \bibinfo {author} {\bibfnamefont {David~B}\ \bibnamefont
  {Phillips}}, \bibinfo {author} {\bibfnamefont {Graeme}\ \bibnamefont
  {Whyte}}, \bibinfo {author} {\bibfnamefont {Tim~A}\ \bibnamefont {Birks}},
  \bibinfo {author} {\bibfnamefont {Yves}\ \bibnamefont {Wiaux}}, \ and\
  \bibinfo {author} {\bibfnamefont {Robert~R}\ \bibnamefont {Thomson}},\
  }\bibfield  {title} {\enquote {\bibinfo {title} {Computational optical
  imaging with a photonic lantern},}\ }\href@noop {} {\bibfield  {journal}
  {\bibinfo  {journal} {Nature communications}\ }\textbf {\bibinfo {volume}
  {11}},\ \bibinfo {pages} {5217} (\bibinfo {year} {2020})}\BibitemShut
  {NoStop}%
\bibitem [{\citenamefont {Kang}\ \emph {et~al.}(2023)\citenamefont {Kang},
  \citenamefont {Kwon}, \citenamefont {Lee}, \citenamefont {Kim}, \citenamefont
  {Hong}, \citenamefont {Yoon},\ and\ \citenamefont {Choi}}]{kang2023tracing}%
  \BibitemOpen
  \bibfield  {author} {\bibinfo {author} {\bibfnamefont {Sungsam}\ \bibnamefont
  {Kang}}, \bibinfo {author} {\bibfnamefont {Yongwoo}\ \bibnamefont {Kwon}},
  \bibinfo {author} {\bibfnamefont {Hojun}\ \bibnamefont {Lee}}, \bibinfo
  {author} {\bibfnamefont {Seho}\ \bibnamefont {Kim}}, \bibinfo {author}
  {\bibfnamefont {Jin~Hee}\ \bibnamefont {Hong}}, \bibinfo {author}
  {\bibfnamefont {Seokchan}\ \bibnamefont {Yoon}}, \ and\ \bibinfo {author}
  {\bibfnamefont {Wonshik}\ \bibnamefont {Choi}},\ }\bibfield  {title}
  {\enquote {\bibinfo {title} {Tracing multiple scattering trajectories for
  deep optical imaging in scattering media},}\ }\href@noop {} {\bibfield
  {journal} {\bibinfo  {journal} {arXiv preprint arXiv:2302.09503}\ } (\bibinfo
  {year} {2023})}\BibitemShut {NoStop}%
\end{thebibliography}%

\onecolumngrid
\newpage

\hspace{5.7cm}{\Large {\bf Supplementary Information}}\\

\noindent{\large{\bf \S 1: Optical setup}}\\

\begin{figure}[H]
    \centering
   \includegraphics[width=0.65\textwidth]{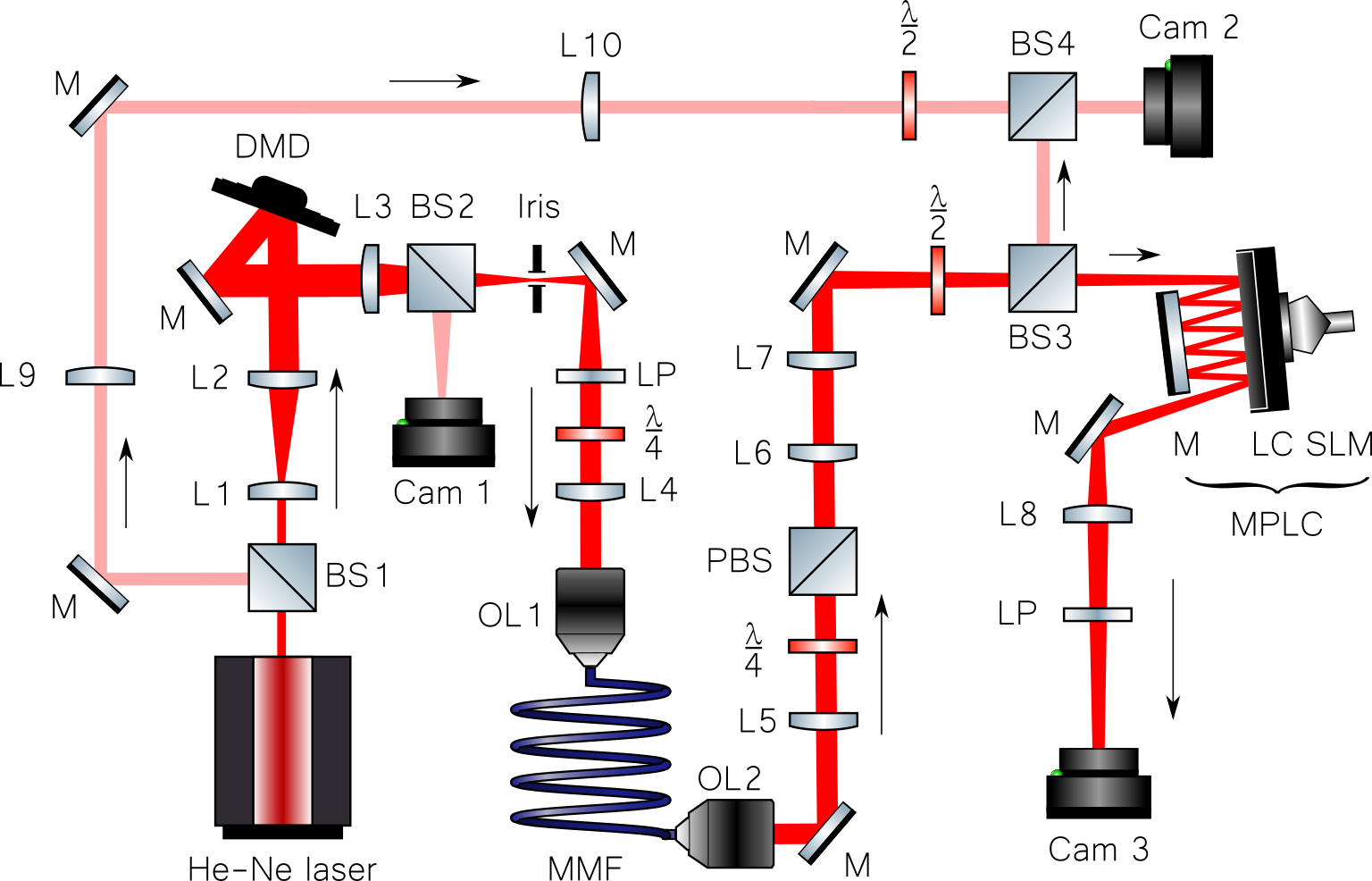}
   \caption{{\bf Schematic of the optical setup used in the experiment.} L -- lens, M -- mirror, BS -- beam splitter, PBS -- polarising beam splitter, Cam -- camera, LP -- linear polariser, OL -- objective lens, DMD -- digital micromirror device, MMF - multimode fibre, LC SLM -- liquid crystal spatial light modulator, $\frac{\lambda}{2}$ and $\frac{\lambda}{4}$ -- half-wave and quarter-wave plates respectively.}
   \label{Fig:suppl_setup}
\end{figure}

Figure~\ref{Fig:suppl_setup} presents a schematic of the optical setup used for implementing the experimental all-optical MMF inverter. A helium-neon laser (Thorlabs HNLS008L-EC) was used as a source of linearly polarised light. After the beam is magnified using a 4-f system of two lenses, a DMD (Vialux V-7001) is used to generate, in its Fourier plane, inputs to the MMF. Camera ``Cam 1" is positioned in the image plane of the fibre input facet (i.e.\ the Fourier plane of the DMD), enabling the intensity profiles of the fields incident on the input of the MMF to be imaged. Prior to entering the fibre, the light passes through a quarter wave-plate to change the polarisation of the incoming light from linear to circular. The optical field generated in the Fourier plane of the DMD is demagnified to the scale of the fibre core using a 4-f system (lens L4 and objective lens OL1). Circularly polarised light on the output of the MMF is converted back to the linear orientation, one component of which is discarded using a polarising beam-splitter. The light emanating from the output facet of the fibre is re-imaged into the plane of the camera "Cam 2", where its intensity can be directly measured. The optical phase of the transmitted fields is retrieved by interfering them with the reference beam and then using off-axis digital holography. The MMF output fields are also re-imaged onto the first plane of the MPLC-based optical inverter. The MPLC is constructed from a liquid crystal SLM (Hamamatsu X13138-01) and a mirror parallel to its screen. After the light is modulated by $M=5$ planes of the MPLC, it is Fourier transformed by a lens and any unmodulated light is discarded with a linear polariser.

\newpage

\noindent{\large{\bf \S 2: Adaptive correction of the optical inverter: MMF bend configurations}}\\

\begin{figure}[H]
    \centering
   \includegraphics[width=0.7\textwidth]{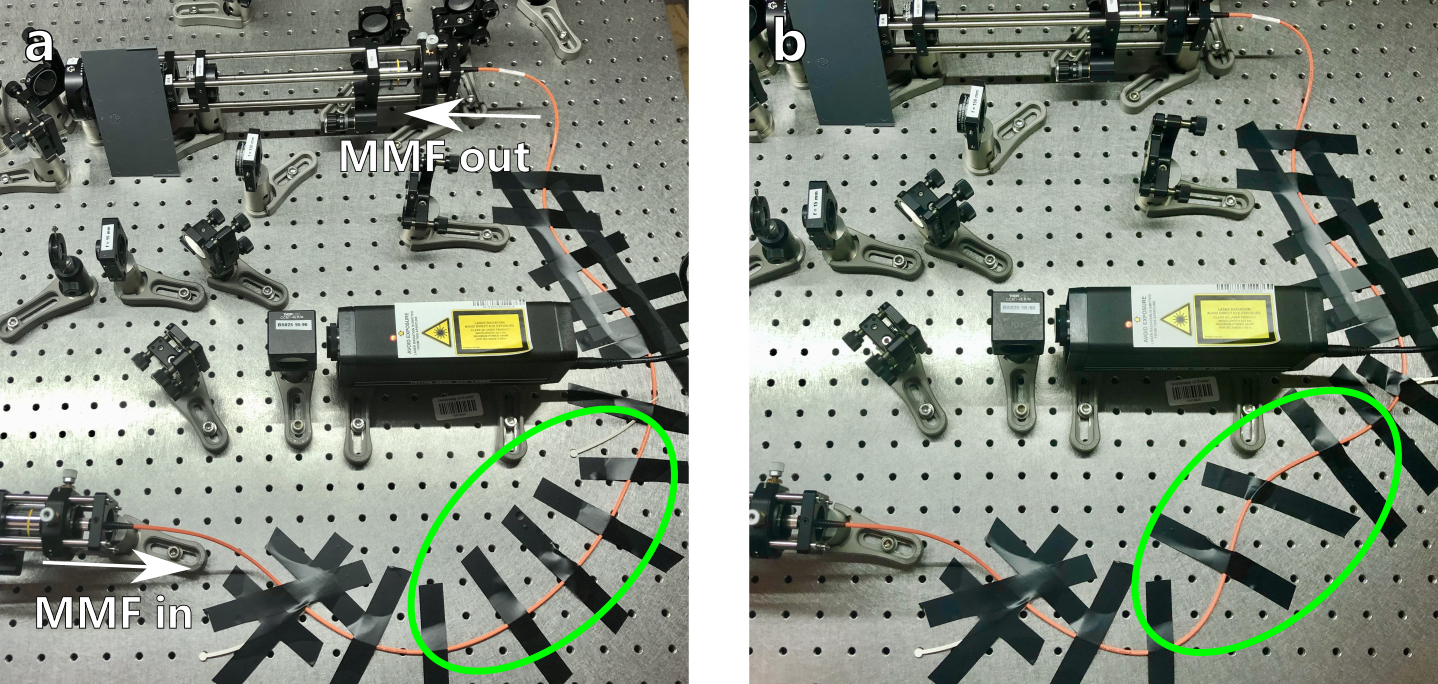}
   \caption{{\bf MMF configuration}. Photographs of the experimental configuration of the multimode fibre (orange jacket) before (a) and after (b) additional bending has been applied to it.}
   \label{Fig:suppl_MMF_bending}
\end{figure}

Figure~\ref{Fig:suppl_MMF_bending} shows the MMF configurations before (a) and after (b) an additional bending has been applied to it. The fibre itself in the presented pictures has an orange jacket and is taped to the optical table to ensure the stability of its TM. The bending configuration states of the fibre correspond to the measurements shown in Fig.~\ref{Fig:exp_inv_bend} for the 'aligned' and the 'perturbed' cases respectively. 

\vspace{1cm}

\noindent{\large{\bf \S 3: Comparison of our gradient ascent algorithm to the wavefront matching method for optical inverter design}}\\

\begin{figure}[H]
    \centering
   \includegraphics[width=0.95\textwidth]{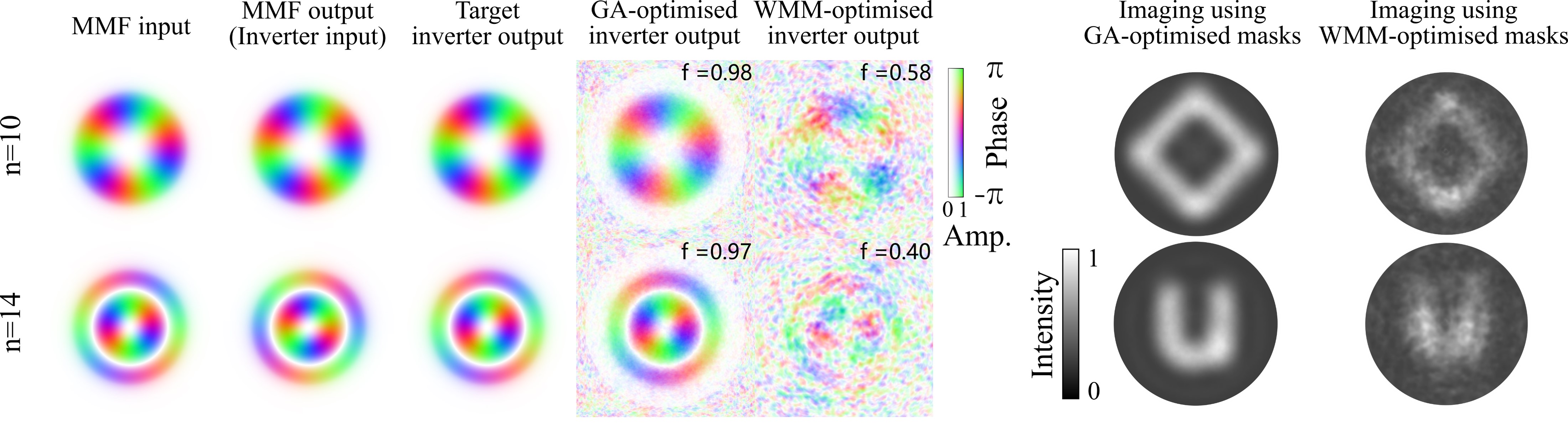}
   \caption{{\bf WMM vs. GA inverse design}. Selected output eigenmodes $n=10$ and $n=14$ of the  visualised alongside the outputs of a system optimised using the wavefront matching method. Examples of the incoherent imaging using the GA- and WMM-optimised inverters are shown as two columns on the right.}
   \label{Fig:suppl_eigenmode_fields_GA_WMM}
\end{figure}

Figure~\ref{Fig:suppl_eigenmode_fields_GA_WMM} compares the simulated performance of two phase mask optimisation techniques that may be used: our gradient ascent algorithm (GA) and the conventional wavefront matching method (WMM). Both algorithms are tasked with designing a 42-mode eigenmode-based optical inverter using 5 phase planes. The quality of two selected output modes are shown, and visually we see the fidelity of the outputs modes from the inverter designed using the WMM is substantially lower than those generated by our GA algorithm. Quantitatively, the fidelity of the first spatial mode (top row) is $f_{n=10}^{\rm GA} = 98\%$, versus $f_{n=10}^{\rm WMM} = 58\%$. For the second example (bottom row), $f_{n=14}^{\rm GA} = 97\%$, and $f_{n=14}^{\rm WMM} = 40\%$. The overall fidelity of the transformation averaged over all the modes of a set for GA is $f^{\rm GA}=97\%$ and for WMM is $f^{\rm WMM}=49\%$, while the simulated average efficiencies (i.e.\ input light utilisation ratios) are $e^{\rm GA}=13\%$, $e^{\rm WMM}=30\%$ (See also Table~\ref{tab:my_label}). Example simulated images using the GA-optimsed inverter and the WMM-optimised inverter are shown on the right hand side of Fig.~\ref{Fig:suppl_eigenmode_fields_GA_WMM}.

\newpage

\noindent{\large{\bf \S 4: Eigenmode-based optical inverter: transformation of all eigenmodes}}\\ 

\begin{figure}[H]
    \centering
   \includegraphics[width=1\textwidth]{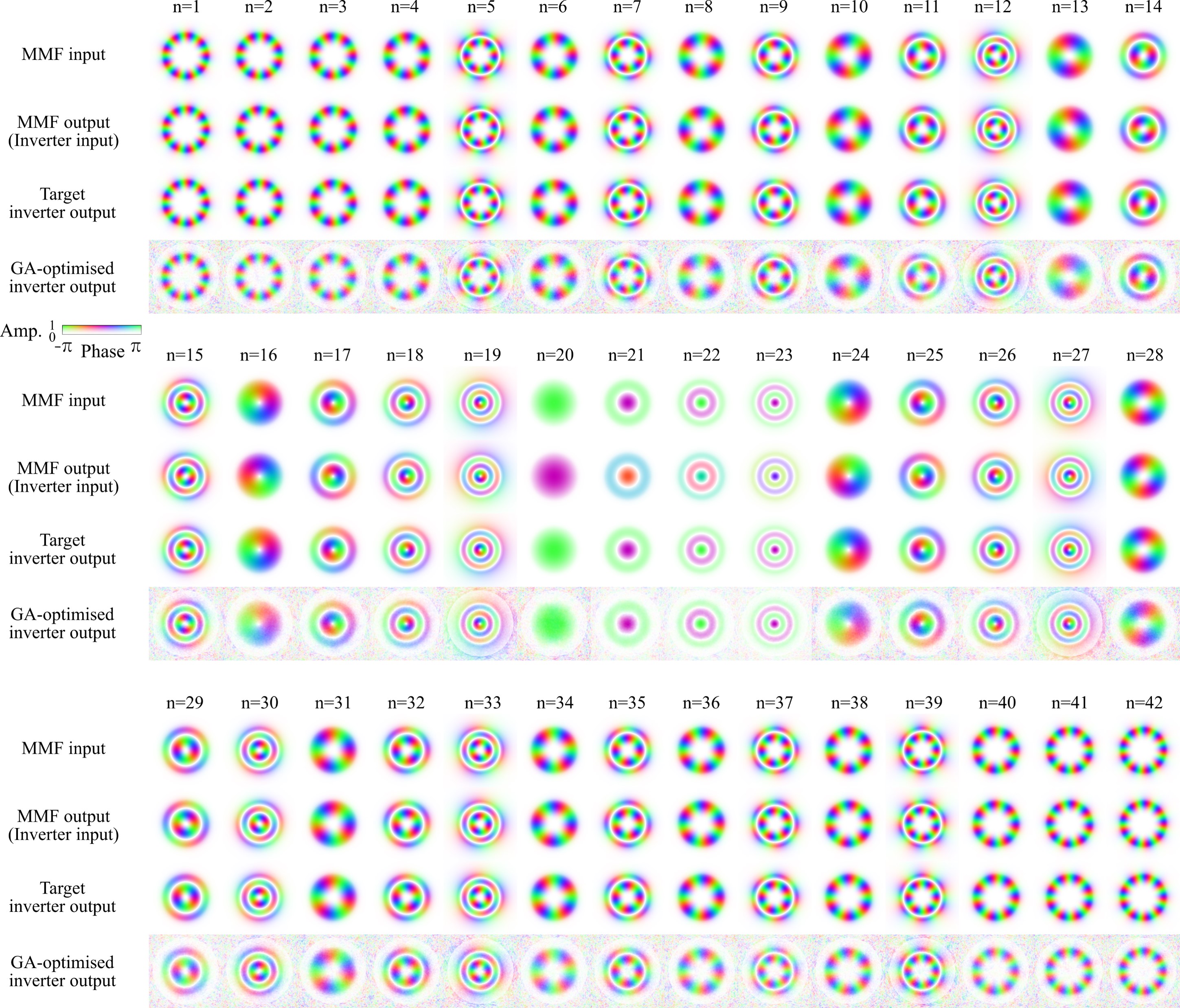}
   \caption{{\bf Visualisation of fields propagating through eigenmode-based optical inverter}. Sets of the $N=42$ PIMs used to design the eigenmode-based optical inverter.}
   \label{Fig:suppl_eigenmode_fields}
\end{figure}

Figure~\ref{Fig:suppl_eigenmode_fields} presents the set of all $N=42$ spatial mode-pairs used to design the eigenmode-based optical inverter matched to an ideal MMF. The MMF eigenmodes accumulate a mode-dependant phase delay $\beta_n L$ as they propagate through the fibre. The multi-plane light conversion system used as the inverter is then designed to act on all $N=42$ output modes simultaneously to compensate for these global phase delays with $M=5$ planes. The fourth row of each mode block of the presented fields demonstrates the spatial modes generated on the output of such an inverter system optimised using our gradient ascent algorithm ($\alpha=1,\gamma=2$). The region of the output plane in which light is allowed to be randomly scattered, in order to enhance output fidelity within the central disk area, can be clearly seen in the outputs.

\newpage

\noindent{\large{\bf \S 5: SVD-based optical inverter: transformation of all modes}}\\

\begin{figure}[H]
    \centering
   \includegraphics[width=1\textwidth]{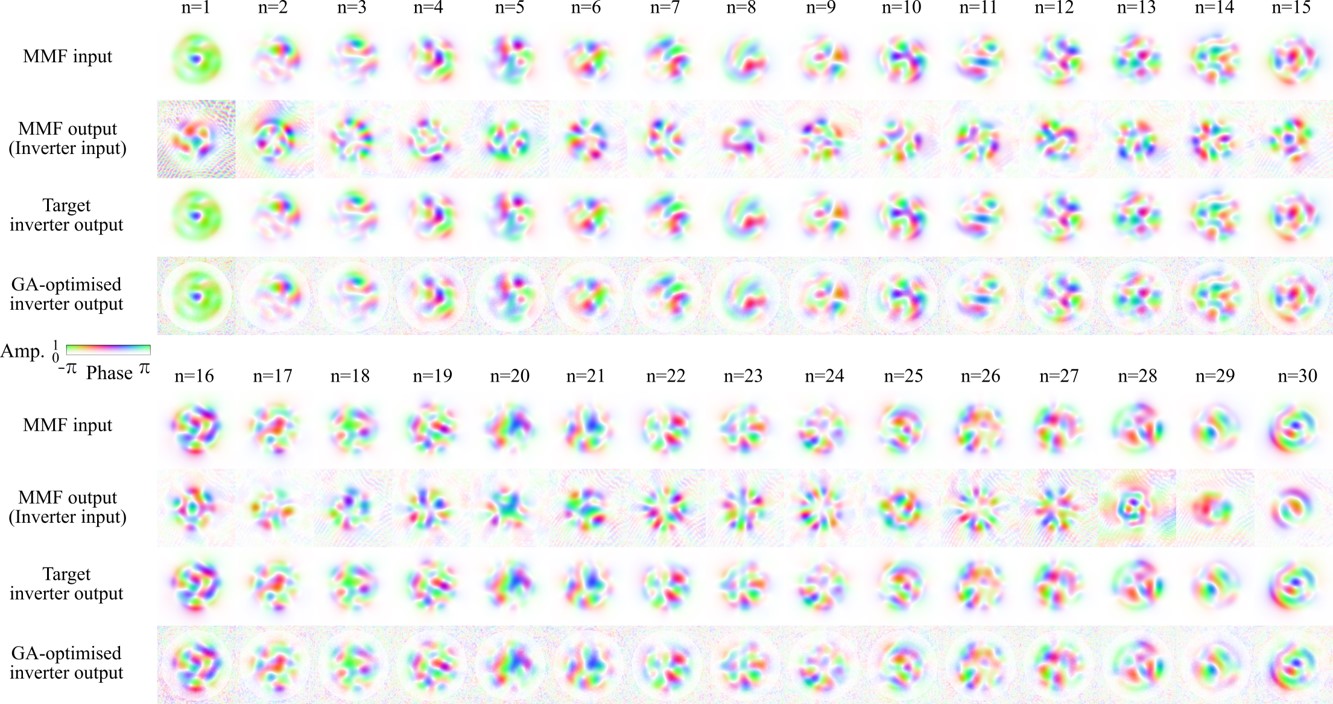}
   \caption{{\bf Visualisation of fields propagating through SVD-based optical inverter}. Sets of $N=30$ optical fields used to design the SVD-based optical inverter, derived from the experimentally measured TM of a multimode fibre.}
   \label{Fig:suppl_SVD_fields}
\end{figure}

Figure~\ref{Fig:suppl_SVD_fields} presents sets of all $N=30$ spatial mode-pairs used to optimse the SVD-based optical inverter, designed to matched to a real MMF. The mode sets are derived from the experimentally measured TM of a MMF. The MMF inputs in the form of $N$ right-hand singular vectors are transformed into $N$ left-hand singular vectors as they propagate through the fibre. The multi-plane light conversion system used as the inverter is then designed to act on all $N=30$ modes simultaneously, to transform left-hand singular vectors back into the right-hand ones with $M=5$ planes. The fourth row of each block of the presented fields demonstrates the spatial modes on the output of such an inverter system optimised using our gradient ascent algorithm ($\alpha=1,\gamma=2$). Once again, the area of the output plane in which light is allowed to be randomly scattered in order to enhance output fidelity within the central disk area, can be clearly seen in the outputs.

\newpage

\noindent{\large{\bf \S 6: Spectral bandwidth of the optical inverters}}\\

\begin{figure}[H]
    \centering
   \includegraphics[width=0.6\textwidth]{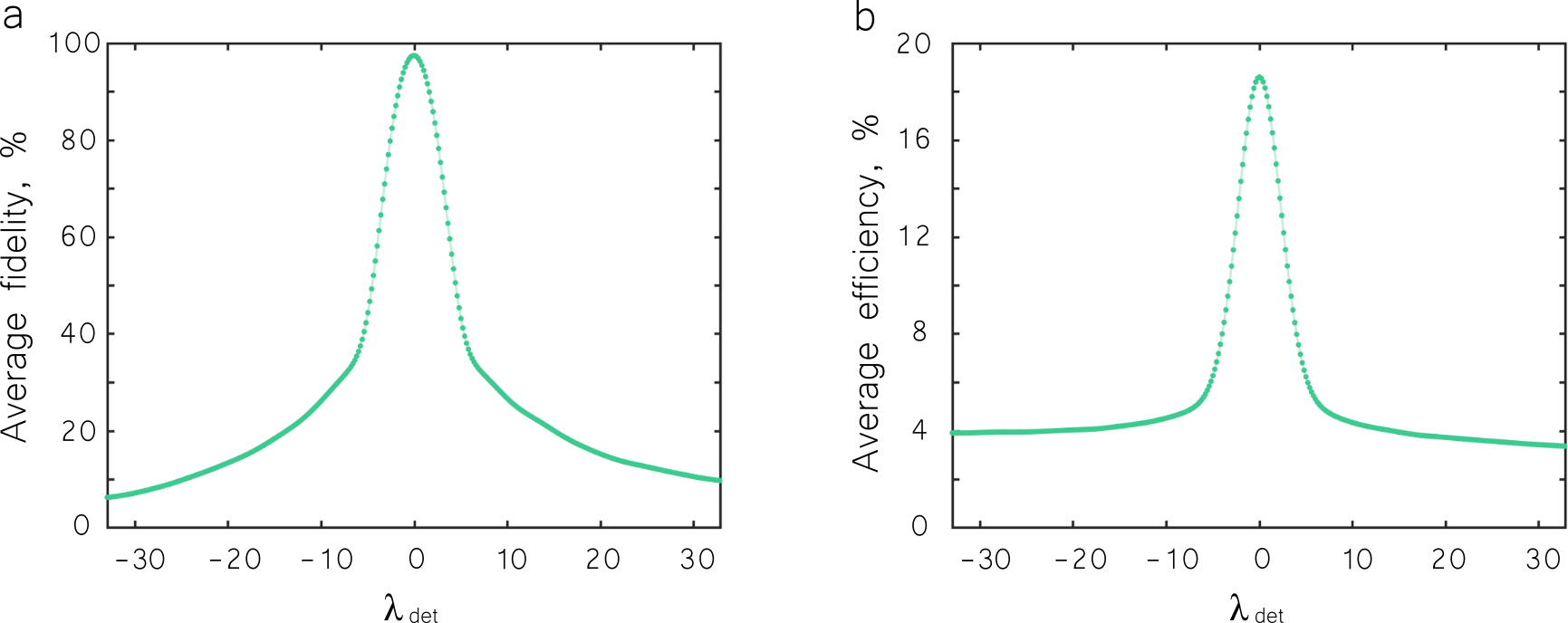}
   \caption{{\bf Simulated spectral bandwidth of the 30-mode SVD-based optical inverter}. The average fidelity (a) and the average efficiency (b) of such a system are plotted as functions of the detuned wavelength.}
   \label{Fig:suppl_bandwidth_SVD}
\end{figure}

Spectral response curves of the simulated optical inverter based on the SVD approach that inverts the transmission matrix of the experimentally measured fibre are shown in Figure~\ref{Fig:suppl_bandwidth_SVD}. These are numerically calculated as follows. First, a set of MPLC $M=5$ phase masks corresponding to the needed transformation is optimised using the gradient ascent algorithm with the base values of $\alpha=1,\gamma=2$ at the original wavelength $\lambda_0 = 633~nm$. After this, the wavelength of the input to the MPLC light is detuned ($\lambda_{\rm det}$) and the phase delays introduced by the phase masks are modified accordingly by multiplication by a factor $\lambda_0/ \lambda_{\rm det}$. After this, the performance of such a mode transformer can be quantified by simulating the propagation of the frequency shifted modes through the device, and calculating the average fidelity and the average efficiency parameters defined in the Methods section of the main text.

The spectral bandwidth of the inverter can then be calculated as the full width half maximum (FWHM) of the average fidelity curve shown in Fig.~\ref{Fig:suppl_bandwidth_SVD}(a). As the minimum level of the average fidelity corresponding to the satisfactory inversion quality is not strictly defined, the spectral bandwidth of such a system may be estimated as $\Delta \lambda = 5-10~nm$. However note that, according to the main text of the paper, the bandwidth of the whole MMF-inverter system is rather limited by the bandwidth of the MMF itself in this case.

\begin{figure}[H]
    \centering
   \includegraphics[width=0.7\textwidth]{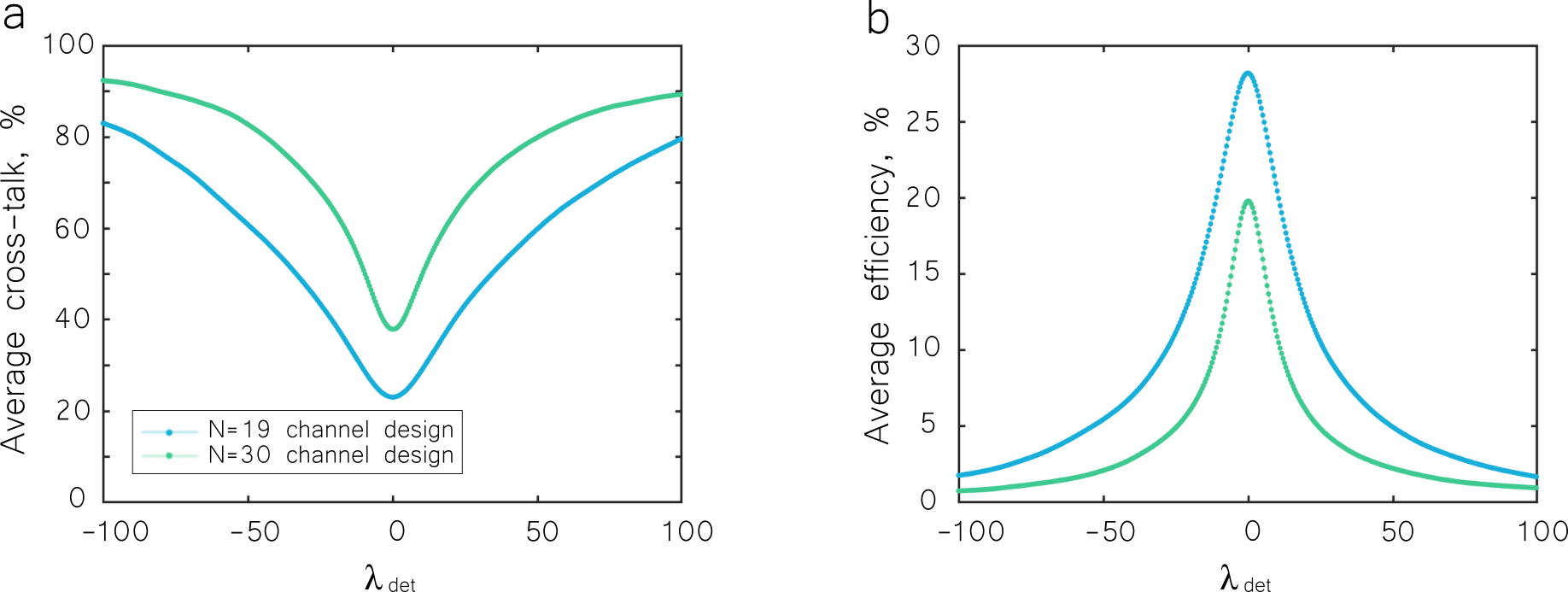}
   \caption{{\bf Simulated spectral response of the MPLC system designed to experimentally operate on $N=19$ (blue) and $N=30$ modes (green) using $M=5$ phase planes in order to invert the scrambling introduced by the fibre.} The average cross-talk (a) and the average efficiency (b) of such systems are plotted as functions of the detuned wavelength.}
   \label{Fig:suppl_bandwidth_exp}
\end{figure}

Figure~\ref{Fig:suppl_bandwidth_exp} presents the simulated spectral response curves of the experimentally realised MMF inverters using the $N=19$ (blue) and $N=30$ speckle mode sorters (green). The average cross-talk and the average efficiency as functions of the detuned wavelength are calculated in the same way as described above for the central wavelength of $\lambda_0 = 633~nm$. We observe a broader FWHM, if compared to Fig.~\ref{Fig:suppl_bandwidth_SVD}. This is due to the fact that the phase masks of these systems were optimised using the WMM, which is equivalent to the gradient ascent optimisation with $\alpha=1, \gamma=0$. This objective function produces less intricate phase masks containing fewer phase wraps thus boosting the spectral bandwidth of the device. The phase masks used were also additionally smoothed to be faithfully represented by the LC SLM screen at the expense of the overall performance~\cite{kupianskyi2023high}. 

\newpage

\noindent{\large{\bf \S 7: Summary of optical inverter performance metrics}}\\

\begin{table}[H]
    \centering
    \begin{tabular}{|c|c|c|c|c|c|}
    \hline
       {\bf Inverter design} & {\bf No.\ of modes} & {\bf No.\ of planes} & {\bf Efficiency (\%)} & {\bf Fidelity (\%)} & {\bf Cross-talk (\%)} \\
       \hline
       Eigenmode (GA) (sim.) & 42 & 5 & 13 & 97 & -\\
       \hline
       Eigenmode (WMM) (sim.) & 42 & 5 & 30 & 49 & -\\
       \hline
       SVD (sim.) & 30 & 5 & 18 & 97 & -\\
       \hline
       speckle sorter (sim.) & 19 & 5 & 27 & - & 24\\
       \hline
       speckle sorter (sim.) & 30 & 5 & 19 & - & 37\\
       \hline
       speckle sorter (exp.) & 19 & 5 & - & - & 33\\
       \hline
       speckle sorter (exp.) & 30 & 5 & - & - & 48\\
       \hline
    \end{tabular}
    \caption{Performance metric summary for all simulated and experimentally implemented optical inverter designs.}
    \label{tab:my_label}
\end{table}

\noindent{\large{\bf \S 8: Description of supplementary movies}}\\

\begin{itemize}
    \item Supplementary movie 1: A simulation of the spatial transformation undergone by input speckle fields as they propagate through the phase planes of a speckle mode-sorter inverter design.    
    \item Supplementary movie 2: The 19-mode experimentally implemented optical inverter outputs as all input channels are sequentially excited. Left panel: Intensity of field input into MMF. Middle panel: Experimentally measured optical field at output of MMF. RIght panel: Intensity of field at output of optical inverter.
    \item Supplementary movie 3: The 30-mode experimentally implemented optical inverter outputs as all input channels are sequentially excited. Left panel: Intensity of field input into MMF. Middle panel: Experimentally measured optical field at output of MMF. RIght panel: Intensity of field at output of optical inverter.
    \end{itemize}

\end{document}